\documentclass{aa}  

\usepackage{graphicx}
\usepackage{booktabs}
\usepackage{multirow}
\usepackage{xcolor}
\usepackage{stfloats}
\usepackage{txfonts}
\usepackage{natbib}
\usepackage{xcolor}
\usepackage[colorlinks=true,citecolor=blue,linkcolor=blue,urlcolor=blue,anchorcolor=blue]{hyperref}
\usepackage[all]{hypcap}

\begin{document}

   \title{The role of velocity dispersion in the Kennicutt-Schmidt relation}


\author{Chryssi Koukouraki
		\inst{1,2}\corrauth{ckoukouraki@physics.uoc.gr}\email{ckoukouraki@physics.uoc.gr}
		\and
		Konstantinos Tassis\inst{1,2}\email{tassis@physics.uoc.gr}
		}	
	\institute{
		University of Crete, Department of Physics \& Institute of
		Theoretical \& Computational Physics, Voutes Campus, 70013 Herakleio, Greece
		\and 
		Institute of Astrophysics,
		Foundation for Research and Technology-Hellas, Plastira 100, Vasilika Vouton, 70013 Heraklion, Greece
		}

 \abstract
   {The resolved Kennicutt-Schmidt (KS) relation is a power-law relation between the surface densities of the star formation rate ($\Sigma_\mathrm{SFR}$) and molecular gas ($\Sigma_\mathrm{mol}$) in star-forming galaxies on kiloparsec scales. Despite its apparent simplicity, it exhibits substantial scatter, suggesting the influence of additional physical parameters beyond gas surface density.}
   {We investigate whether molecular gas velocity dispersion ($\Delta v$) acts as a hidden parameter in the resolved KS relation in nearby galaxies.}
   {We use spatially resolved measurements of $\Sigma_\mathrm{SFR}$, $\Sigma_\mathrm{mol}$, and $\Delta v$ from the ALMaQUEST, EDGE–CALIFA, and PHANGS–ALMA surveys and from observations of the single galaxy M51. We identify two groups of galaxies based on the direction of the $\Delta v$ gradient with respect to the best-fit KS line: one in which the $\Delta v$ gradient is completely aligned with the KS line (group A) and thus does not contribute to the scatter, and one in which it has a component perpendicular to the line (group B) and contributes to the scatter. For each group, we fit the KS relation and examine the dependence of the residuals ($Res$) of the fit on $\Delta v$. }
   {In group B, we find a weak but statistically significant correlation between $Res$ and $\Delta v$, which is consistent across all surveys.
   A three-dimensional fit including $\Delta v$ yields $\Sigma_\mathrm{SFR} \propto \Sigma_\mathrm{mol}^{1.15} \Delta v^{-0.59}$.
   } 
   {There is a set of star-forming galaxies for which the velocity dispersion acts as a secondary parameter in the KS relation, with higher $\Delta v$ corresponding to lower $\Sigma_\mathrm{SFR}$ at fixed $\Sigma_\mathrm{mol}$. 
   We find no obvious correlation between membership in this set and the other properties examined.}


   \maketitle
   \nolinenumbers
%
\section{Introduction}

Scaling relations in astrophysics serve as fundamental tools in understanding physical processes governing galaxies. Empirical correlations such as the Tully-Fisher relation \citep{Tully1977}, the Faber-Jackson relation \citep{Faber1976}, and the M-$\sigma$ relation \citep{Ferrarese2000} have been crucial for understanding the physical mechanisms underlying galaxy evolution. These relations, though widely used, often lack a robust theoretical explanation. This warrants further examination of their dependence on additional quantities, which may uncover the underlying physical mechanisms behind these empirical relations (e.g., fundamental plane of elliptical galaxies \citep{Djorgovski1987}).

In the context of star formation, one of the most widely studied and used scaling relations is the Kennicutt–Schmidt (KS) relation, which connects the rate at which stars form in galaxies (SFR) to the amount of gas available. \citet{Schmidt1959} first proposed that the SFR volume density is related to the gas volume density through a power-law, i.e., $\rho_\mathrm{SFR} \propto \rho_\mathrm{gas}^n$. \citet{Kennicutt1989,Kennicutt1998} then modified this relation to involve surface densities across entire galaxies instead, demonstrating empirically that $\Sigma_\mathrm{SFR} \propto \Sigma_\mathrm{gas}^n$, where $\Sigma_\mathrm{SFR}$ is the SFR surface density and $\Sigma_\mathrm{gas}$ is the total (atomic + molecular) gas mass surface density.
Subsequent studies have shown that this correlation is significantly tighter when considering only the molecular gas component, $\Sigma_\mathrm{mol}$, which is the direct fuel for star formation, while atomic gas exhibits a much weaker correlation \citep[e.g.,][]{Bigiel2008,Leroy2008,Schruba2011}.
More recent studies have moved beyond global, galaxy-averaged measurements to spatially resolved (kpc or sub-kpc) analyses \citep[e.g.,][]{Kennicutt2007}, showing that the KS relation can vary across different galactic environments and scales, reflecting the local physical conditions of the interstellar medium.

Despite its apparent simplicity, the KS relation exhibits substantial intrinsic scatter, both in its global and resolved forms, often spanning more than an order of magnitude in $\Sigma_\mathrm{SFR}$ at fixed $\Sigma_\mathrm{mol}$, implying that additional physical parameters, beyond gas surface density alone, influence star formation efficiency.
Additionally, the exponent of the KS relation varies significantly among surveys.

Various theoretical models have sought to explain the physical origin of the KS relation. For example, \citet{Silk1997} and \citet{Elmegreen1997} suggest that the efficiency of star formation depends on the galactic dynamical timescale, linking gas consumption to orbital dynamics (``dynamical KS'' or ``Silk–Elmegreen'' law); \citet{Tan2000} proposes that galactic shear-induced cloud collisions enhance star formation efficiency (shear-driven GMC--GMC collision model); \citet{Ostriker2010} suggest that star formation is self-regulated through a balance between the vertical weight of the interstellar medium and the pressure generated by stellar feedback; \citet{Tassis2007} suggests that if the ISM has a multifractal geometry, then the KS relation naturally emerges from its structure.

Observational studies have  also investigated whether additional parameters can account for departures from the KS relation. For example, \citet{Reyes2019} investigated whether secondary parameters such as stellar surface density and dynamical timescale can account for the observed scatter in the KS relation by examining correlations between these quantities and the residuals of the KS relation. More directly related to the present work, \citet{Wang2020} investigated molecular gas velocity dispersion using ALMA CO(1--0) observations and Spitzer-based SFR estimates for approximately 200pc regions in six nearby galaxies. Their modified KS relation indicated lower $\Sigma_{\rm SFR}$ at higher velocity dispersion for a fixed CO luminosity surface density.

In this work, we investigate the influence of molecular gas velocity dispersion, $\Delta v$, on the resolved KS relation, using kiloparsec-scale measurements for a wide sample of nearby galaxies from several surveys. Our analysis reveals a statistically significant, though relatively weak, dependence for a group of galaxies: for a fixed $\Sigma_\mathrm{mol}$, regions with higher $\Delta v$ exhibit systematically lower $\Sigma_\mathrm{SFR}$. This result suggests that $\Delta v$ may act as an additional hidden parameter modulating star formation efficiency.
\section{Data}
\begin{figure}
  \includegraphics[width=\hsize]{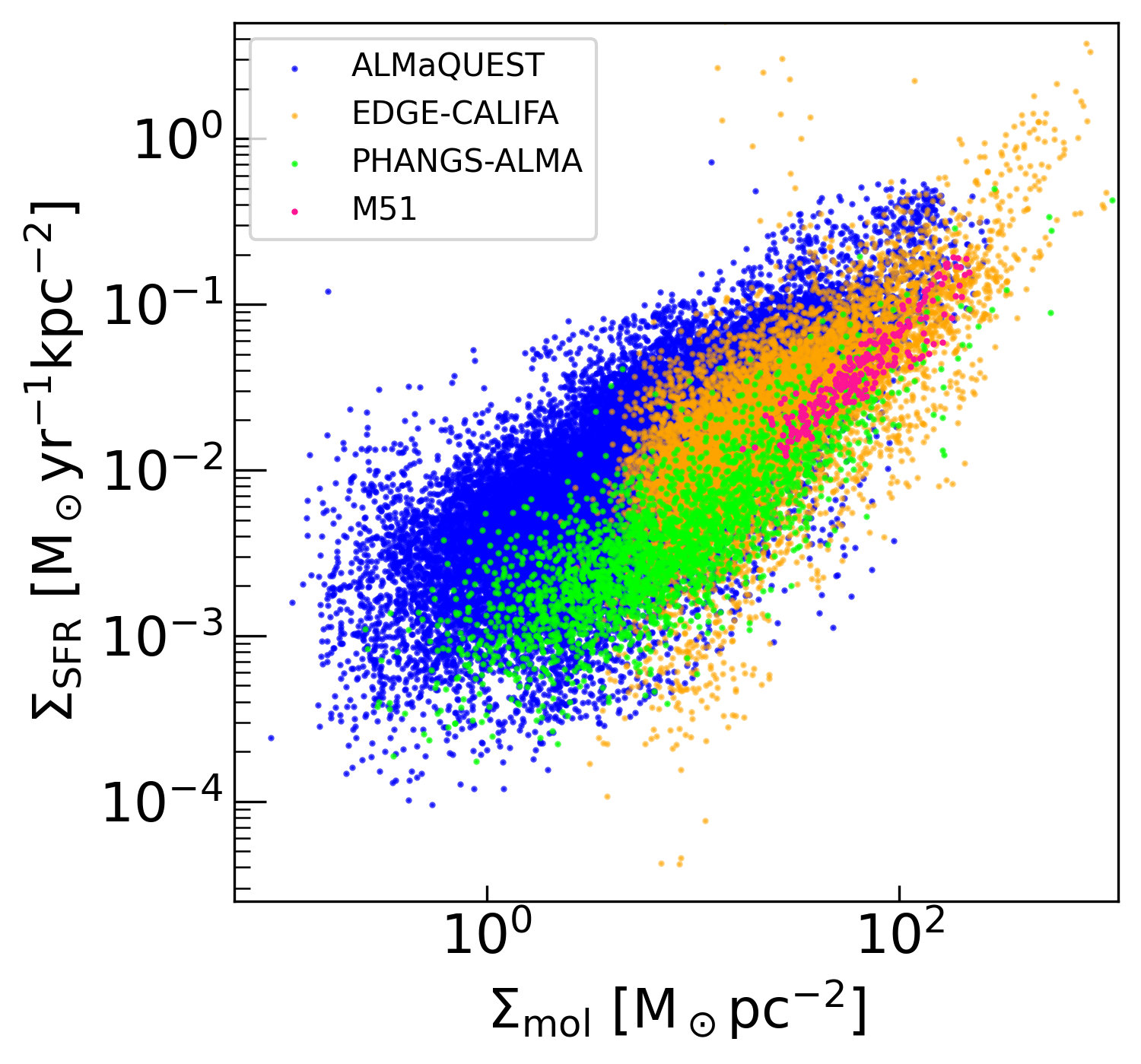}
  \caption{Resolved KS relation of all surveys.}
  \label{KS_all}
\end{figure}

We investigate the role of velocity dispersion, $\Delta v$, in the KS relation using observations of $\Sigma_\mathrm{SFR}$, $\Sigma_\mathrm{mol}$, and $\Delta v$ from the ALMaQUEST \citep{Lin2020}, EDGE--CALIFA \citep{Bolatto2017}, and PHANGS--ALMA \citep{Leroy2021} surveys, as well as the M51 dataset from \citet{Leroy2017}. For each survey, we retain all galaxies for which spatially matched (positive) measurements of the above quantities are available. In the following subsections, we briefly summarize how these quantities are derived in each survey and describe the additional processing steps applied in this work. We refer the reader to the respective survey publications for further details. We present the KS relation of all surveys in Fig.~\ref{KS_all}.

\subsection{ALMaQUEST Survey}
The ALMaQUEST survey combines MaNGA integral-field spectroscopy \citep{manga1,manga2} with ALMA CO(1--0) observations, providing spatially resolved measurements of $\Sigma_{\mathrm{SFR}}$, $\Sigma_{\mathrm{mol}}$, and $\Delta v$ across a sample of nearby galaxies \citep{Lin2019}. 
The MaNGA spectroscopy has a typical spatial resolution of $\sim$2.5$\arcsec$ and a spectral resolution of $\sim$70 km/s, while the ALMA CO maps have a resolution of $\sim$2$\arcsec$ ($\sim$1 kpc at the median distance of the sample) and a spectral resolution of $\sim$10 km/s.
Values of $\Sigma_{\mathrm{SFR}}$ were derived from MaNGA data by converting extinction-corrected H$\mathrm{\alpha}$ fluxes (corrected using the Balmer decrement) into SFR following \citet{Kennicutt1998}, and then dividing by the physical area of each spaxel. The resulting values are also corrected for inclination using the axial ratios from the NASA Sloan Atlas (NSA) catalog.
$\Sigma_{\mathrm{mol}}$ is computed from ALMA CO(1--0) integrated intensity maps by applying a constant CO--to--H$_2$ conversion factor [$\alpha_\mathrm{CO} = 4.35 \mathrm{M_{\odot}}$ (K km s$^{-1}$ pc$^{2}$)$^{-1}$], and dividing the resulting molecular gas mass by the spaxel area, with spatial resolution matched to MaNGA.
The velocity dispersion $\Delta v$ is obtained from the second-moment maps of the ALMA CO(1--0) data cubes, representing the intensity-weighted velocity dispersion within an appropriately selected velocity range.

Of the 46 ALMaQUEST galaxies, 45 are included in the analysis. We restrict our analysis to spaxels with $\Delta v>5$ km/s, corresponding to half the spectral channel width. At lower values, the observed linewidths approach the instrumental resolution and cannot be robustly separated from instrumental broadening, leading to large uncertainties in the inferred velocity dispersion \citep[e.g.,][]{Leroy2016,Sun2018}.

\begin{figure}
  \includegraphics[width=\hsize]{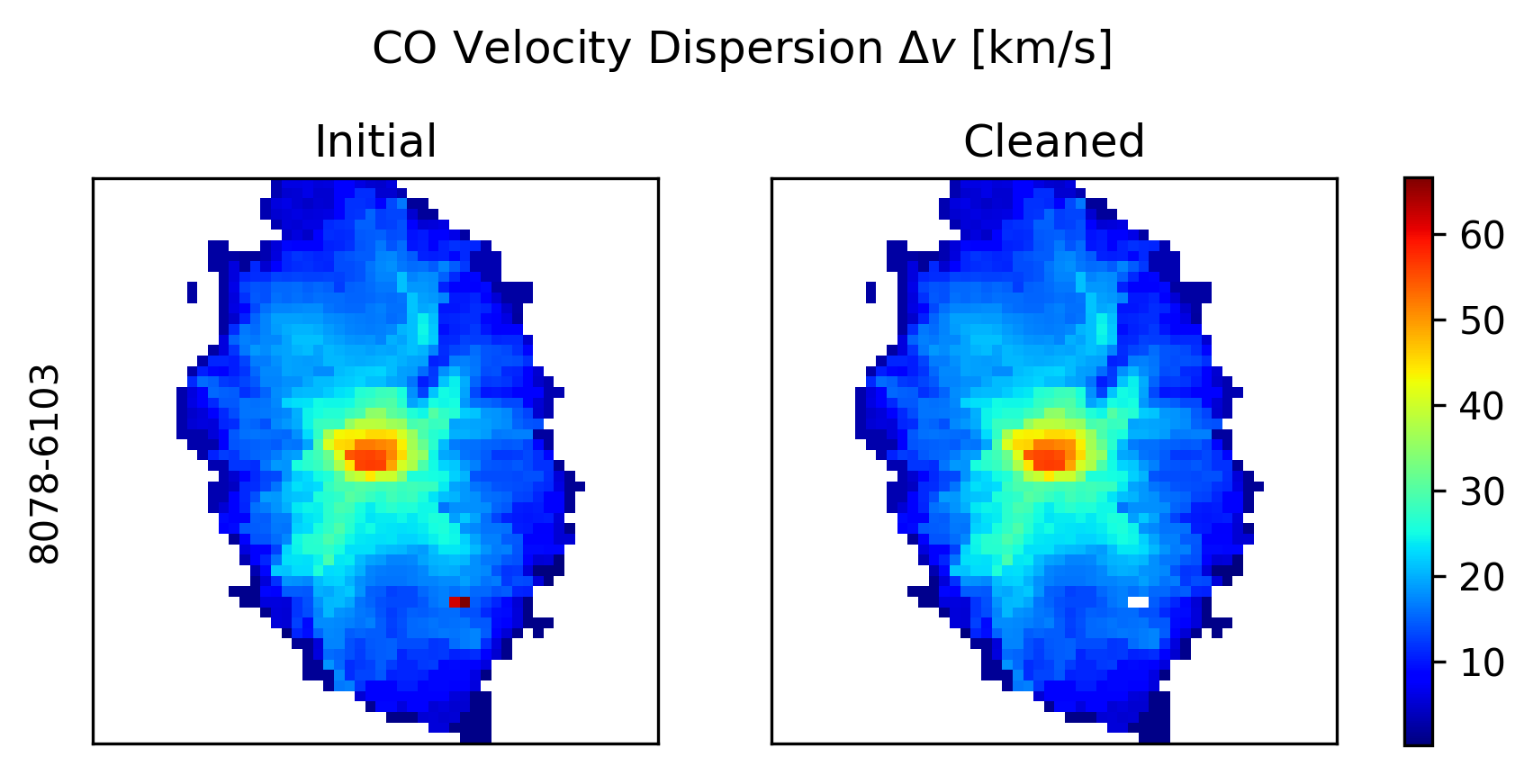}
  \caption{Example of a velocity dispersion map before and after removing noise spikes.}
  \label{alma_cleaning_example}
\end{figure}

To remove isolated noise spikes from the $\Delta v$ maps, we smoothed each $\Delta v$ map with a Gaussian kernel ($\sigma=1$ pixel), and identified and removed outliers via $8\sigma$ clipping of the residuals. To preserve the galaxy centers, all flagged pixels within a 7-pixel radius around the geometric center of the map were restored to their original values. We present an example of a galaxy before and after the removal of noise spikes in Fig.~\ref{alma_cleaning_example}.

We also correct $\Sigma_\mathrm{mol}$ and $\Delta v$ for inclination effects ($\Sigma_\mathrm{SFR}$ was already corrected), following \citet{Sun2022}. Specifically, the reported $\Sigma_\mathrm{mol}$ and $\Delta v$ values are multiplied by $\cos i$ and $(\cos i)^{0.5}$, respectively, where the inclination, $i$, of each galaxy was derived from the axial ratios listed in Table 1 of \citet{Ellison}.

\subsection{PHANGS--ALMA Survey}

$\Sigma_\mathrm{SFR}$ was derived from a combination of the Galaxy Evolution Explorer (GALEX) FUV and Wide-field Infrared Survey Explorer (WISE) 22 $\mu$m emission following the calibration of \citet{Leroy2021}, with attenuation-corrected H$\alpha$ + 22 $\mu$m used where PHANGS--H$\alpha$ data were available \citep{Sun2022}. The maps, with native resolutions of $\sim$1 kpc (FUV/IR) or $\sim$15$\arcsec$ (H$\alpha$ + IR), were convolved to a common $\sim$1 kpc beam, projected to face-on geometry, and sampled in 1.5 kpc hexagonal apertures. $\Sigma_\mathrm{mol}$ was obtained from PHANGS–ALMA CO(2--1) maps with a $\sim$150 pc spatial resolution \citep{Leroy2021} using a metallicity-dependent CO--to--H$_2$ conversion factor, and averaged within 1.5 kpc hexagonal apertures. $\Delta v$ (denoted as $\langle \sigma_\mathrm{pix, \ 150 pc} \rangle$ in \citealp{Sun2022}) was obtained from the CO(2--1) second-moment maps at 150 pc resolution and then averaged within 1.5 kpc hexagonal apertures. Both $\Sigma_\mathrm{mol}$ and $\Delta v$ measurements are corrected for inclination effects. Of the 80 PHANGS--ALMA galaxies, 53 are included in the analysis.

\subsection{EDGE--CALIFA Survey}

The EDGE--CALIFA survey combines optical integral-field spectroscopy from CALIFA \citep{califa} with CO(1--0) observations from the CARMA EDGE survey \citep{Bolatto2017} to provide spatially resolved measurements of molecular gas and star formation in nearby galaxies. $\Sigma_\mathrm{SFR}$ is obtained from extinction-corrected H$\alpha$ emission, $\Sigma_\mathrm{mol}$ is derived from CO(1--0) line intensities using a standard CO--to--H$_2$ conversion factor, and $\Delta v$ is measured from the CO second-moment maps. The CO maps have a spatial resolution of $\sim$4$\arcsec$ ($\sim$1.4 kpc at the typical galaxy distance) and a spectral resolution of $\sim$6.5 km/s. We correct $\Sigma_{\mathrm{mol}}$ and $\Delta v$ for inclination following \citet{Sun2022}. Of the 125 EDGE--CALIFA galaxies, 105 are included in the analysis.

\subsection{M51}

$\Sigma_{\mathrm{SFR}}$ is estimated from the total infrared (surface) brightness, which is obtained from Herschel and Spitzer data with a 1.1 kpc resolution (see their Eq. (2)). The Spitzer data refer to 24 µm and 70 µm intensities, and the Herschel data to 160 µm and 250 µm intensities. 
$\Sigma_{\mathrm{mol}}$ is obtained from the CO(1--0) emission line of the PAWS Survey \citep{PAWS}, which has a native resolution of 40 pc and a spectral resolution of $\sim$5 km/s, using a standard CO--to--H$_2$ conversion factor, and then averaged over each observed SFR region of 1.1 kpc.
$\Delta v$ is obtained from the intensity-weighted CO(1--0) RMS linewidths of the PAWS survey, and is similarly averaged over 1.1 kpc regions. Note that each spectrum is recentered around the local mean velocity, thus not accounting for any bulk motions above the resolution limit.
\section{Results}

\begin{figure*}
  \centering
  \includegraphics[width=\hsize]{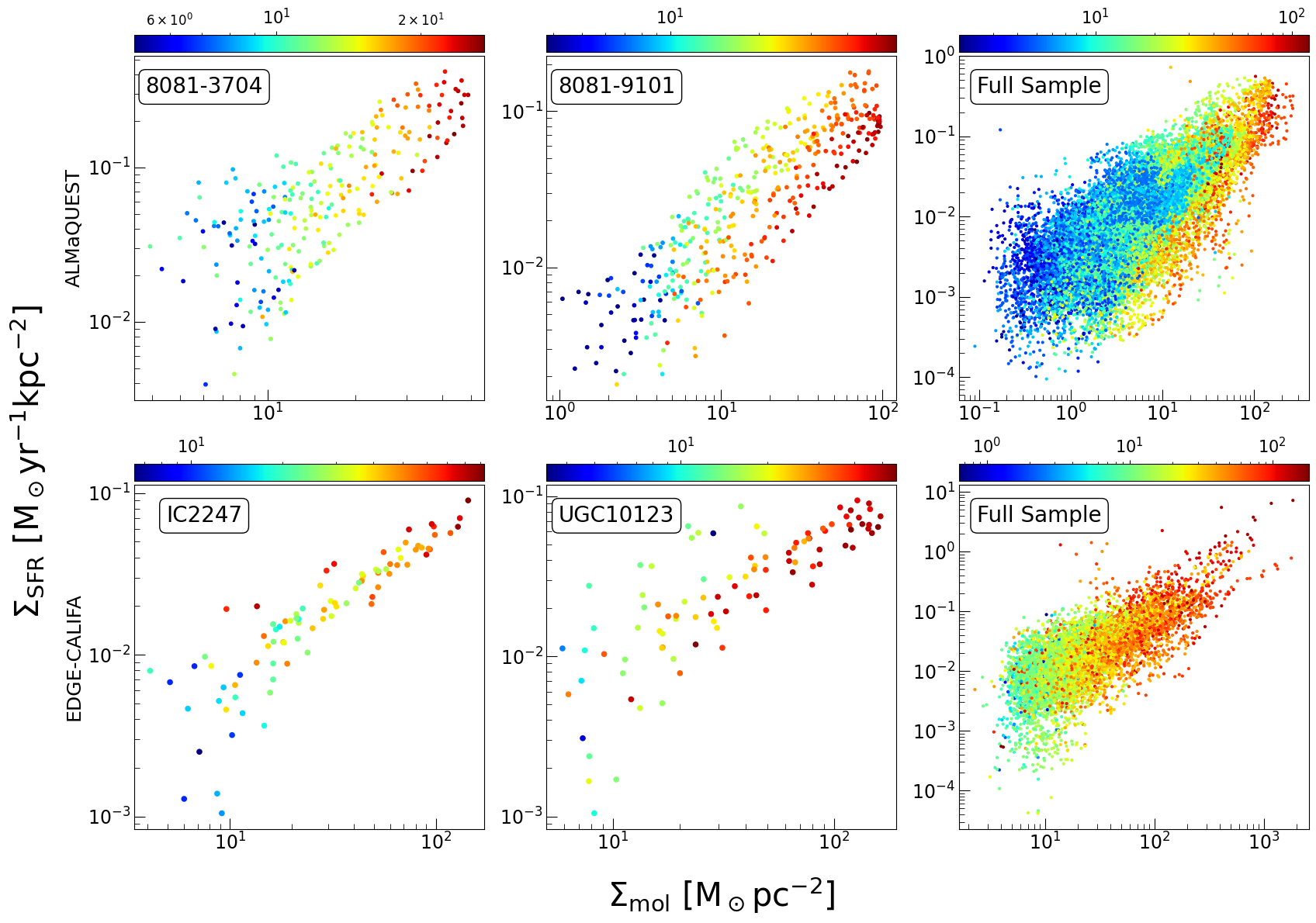}
  \caption{Examples of KS relations of individual galaxies color-coded by $\Delta v$ in the ALMaQUEST and EDGE-CALIFA surveys. Left column: examples of galaxies belonging to group A. Middle column: examples of galaxies belonging to group B. Right column: full samples of galaxies.}
\label{perp_along_all_examples}
\end{figure*}

\begin{figure*}
  \centering
  \includegraphics[width=\hsize]{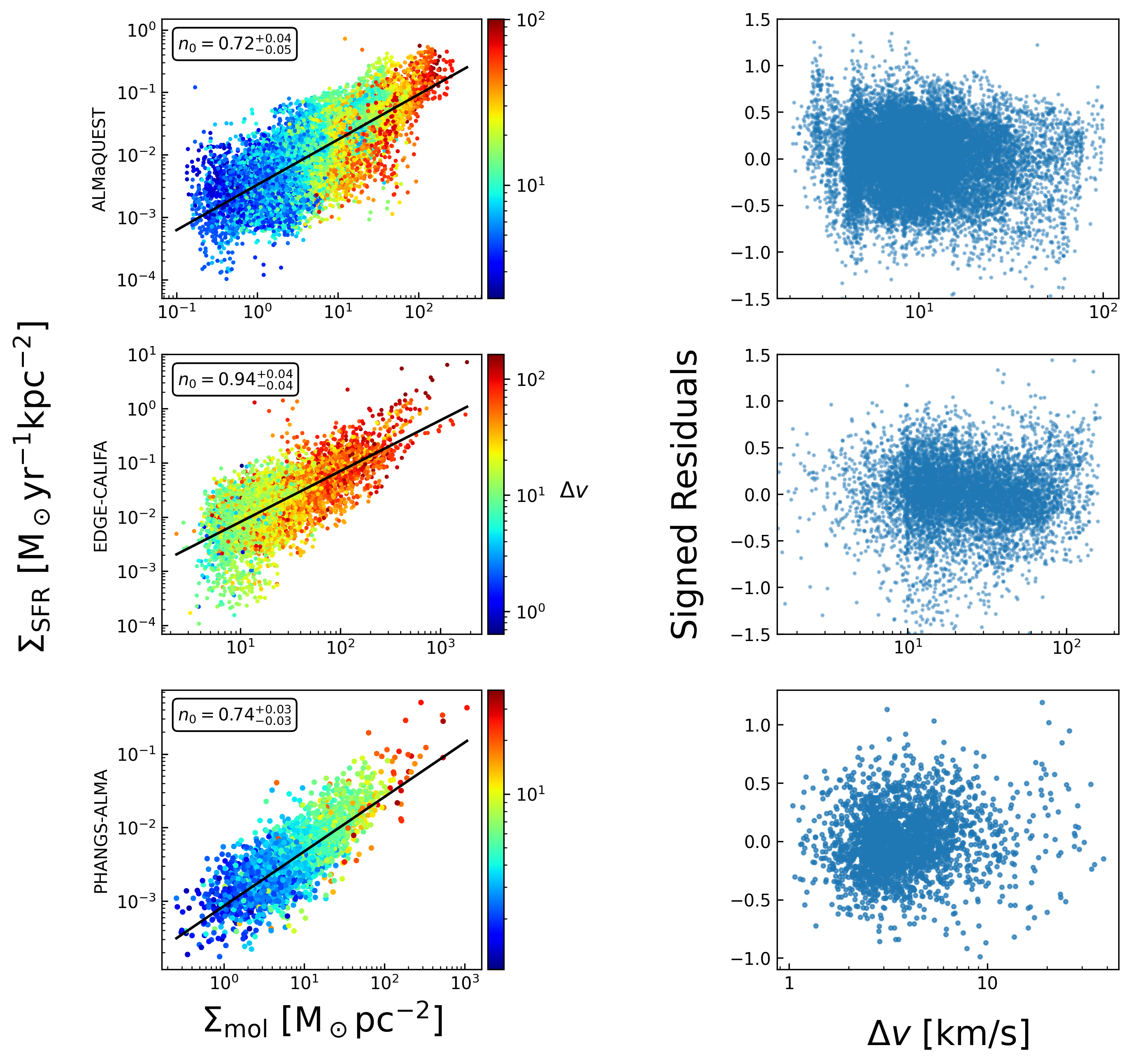}
  \caption{Left column: KS relation for group A galaxies, color-coded by $\Delta v$. The color gradient is along the best-fit KS line.
  Right column: Signed residuals of the best KS fit against $\Delta v$. No significant dependence exists.}
  \label{fig:KS_DV_err_along}
\end{figure*}

\begin{figure*}
  \centering
  \includegraphics[width=\hsize]{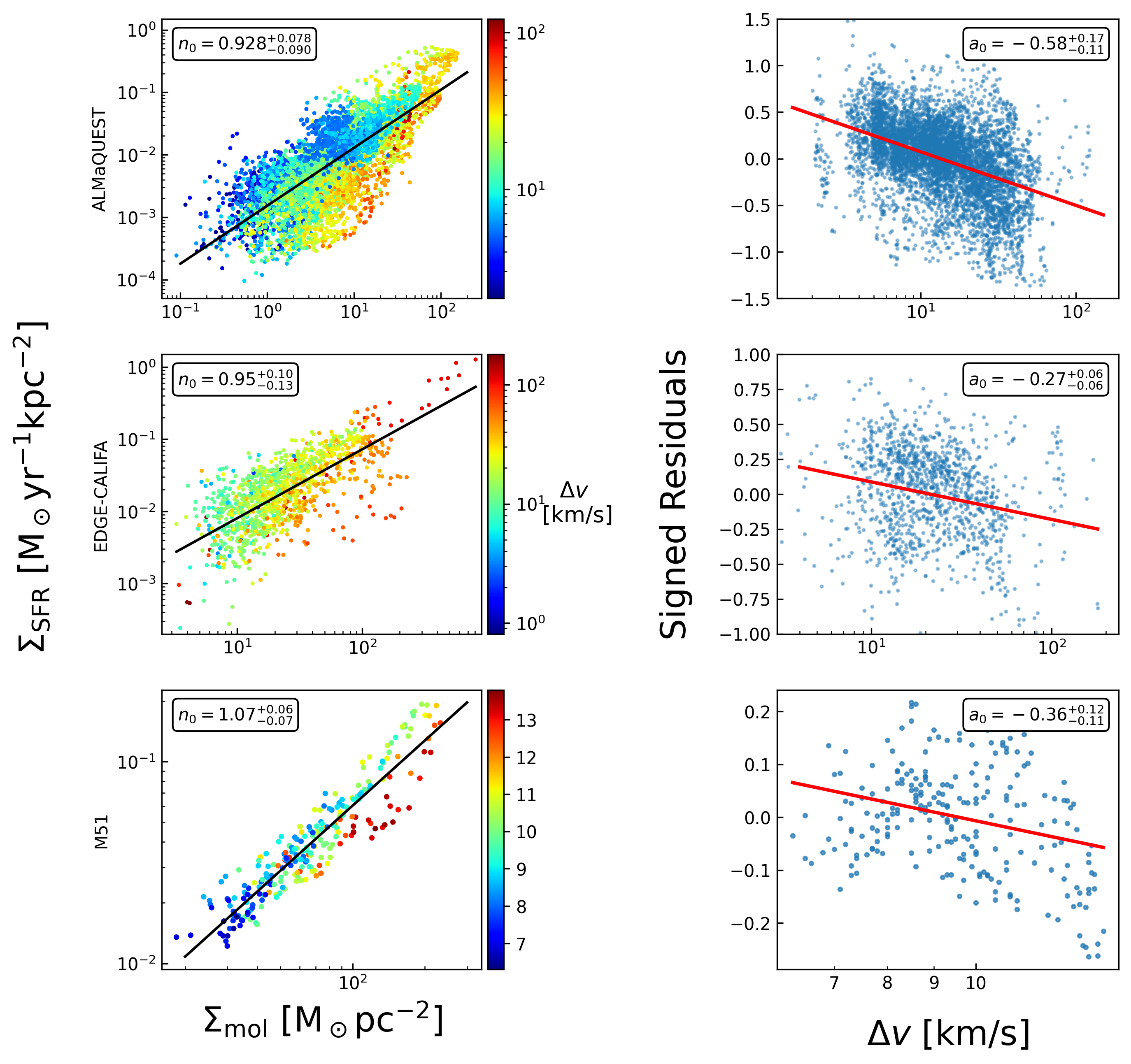}
  \caption{Left column: KS relation for group B galaxies, color-coded by $\Delta v$. The color gradient has a component perpendicular to the best-fit KS line, suggesting that $\Delta v$ is a hidden parameter in the KS relation of those galaxies.
  Right column: Signed residuals of the best KS fit against $\Delta v$. The residuals systematically decrease with increasing $\Delta v$.
  }
  \label{fig:KS_DV_err_perp}
\end{figure*}

For each galaxy in the ALMaQUEST, PHANGS--ALMA, and EDGE--CALIFA surveys, as well as for M51, we make the standard KS plot, color-coding each point by its corresponding $\Delta v$ value. We identify two groups of galaxies in the ALMaQUEST and EDGE--CALIFA samples: group A, where the $\Delta v$ color gradient is entirely along the fitted KS line; and group B, where it also has a component perpendicular to it, indicating that velocity dispersion might be a ``hidden parameter'' in the KS relation of the latter group. In Fig.~\ref{perp_along_all_examples}, we present examples of the KS relation (color-coded by $\Delta v$) of galaxies that belong to group A and group B in the ALMaQUEST and EDGE--CALIFA surveys, as well as the KS relation of the total samples of these surveys. To objectively classify galaxies into group A and group B, we constructed three multiscale diagnostics and fitted an unsupervised Gaussian-mixture model (GMM) to the combined ALMaQUEST and EDGE--CALIFA samples.
When the model fitted to the ALMaQUEST and EDGE--CALIFA galaxies was subsequently applied to the PHANGS--ALMA galaxies and to M51, all PHANGS--ALMA galaxies were independently assigned to group A and M51 was assigned to group B. 
We explain the model in detail and provide the full classification of ALMaQUEST and EDGE--CALIFA galaxies into groups A and B in Appendix~\ref{app:unsupervised-classification}. We attribute the absence of group B galaxies in the PHANGS--ALMA sample, at least in part, to the conservative construction and consequent sparse spatial sampling of the PHANGS--ALMA linewidth maps, which reduces the sensitivity of our classification to a component of the $\Delta v$ gradient perpendicular to the KS relation. We therefore caution against interpreting this absence as conclusive evidence that such behavior is intrinsically absent from the PHANGS--ALMA galaxy population. We discuss this limitation and present example maps in Appendix~\ref{app:phangs}.

For each group within each survey, we combine all regions from all galaxies and perform a KS (least-squares) fit of the form 
\( \log\left( \Sigma_\mathrm{SFR} \right) = n_0\log\left( \Sigma_\mathrm{mol} \right) + C_0 \) (see Figs.~\ref{fig:KS_DV_err_along} and~\ref{fig:KS_DV_err_perp}, left column). We do not account for the intrinsic errors in the
measurements or the uncertainties associated with the conversion of observables to physical quantities. To account for the non-independence of spatial regions belonging to the same galaxy, we estimate the uncertainties of the fitted parameters using a galaxy-level bootstrap. In each of $10^4$ bootstrap realizations, we draw, with replacement, the same number of galaxies as in the corresponding survey and group, include all spatial regions associated with each selected galaxy, and repeat the fit. The quoted coefficients correspond to the fit to the original sample, while their lower and upper uncertainties are determined from the 16th and 84th percentiles of the bootstrap distributions (these intervals are approximately equivalent to $1\sigma$, since the distributions are close to Gaussian). For M51 (which is a single galaxy), we perform similar spatial block bootstraps. We present the best-fit coefficients, as well as the dispersion in $\log\Sigma_\mathrm{SFR}$ ($\sigma$) of the fit in Table~\ref{tab:KS_all_fit_params}. 
\begin{table*}
\centering
\caption{Fitted parameters of the KS relation of the two groups of galaxies}
\label{tab:KS_all_fit_params}

\renewcommand{\arraystretch}{2.0}

\begin{tabular}{ccccc}
\toprule
Survey
& Group
& $n_0$
& $C_0$
& $\sigma$ \\
\midrule

\multirow[c]{2}{*}[-0.4ex]{ALMaQUEST}
& A
& $0.724\mkern4mu{}^{+0.040}_{-0.049}$
& $-2.483\mkern4mu{}^{+0.057}_{-0.048}$
& $0.341$ \\

& B
& $0.928\mkern4mu{}^{+0.078}_{-0.090}$
& $-2.816\mkern4mu{}^{+0.095}_{-0.113}$
& $0.398$ \\

\addlinespace[0.2em]
\midrule
\addlinespace[0.2em]

\multirow[c]{2}{*}[-0.4ex]{EDGE--CALIFA}
& A
& $0.936\mkern4mu{}^{+0.038}_{-0.038}$
& $-3.026\mkern4mu{}^{+0.059}_{-0.057}$
& $0.368$ \\

& B
& $0.955\mkern4mu{}^{+0.096}_{-0.135}$
& $-3.050\mkern4mu{}^{+0.191}_{-0.135}$
& $0.352$ \\

\addlinespace[0.2em]
\midrule
\addlinespace[0.2em]

PHANGS--ALMA
& A
& $0.741\mkern4mu{}^{+0.026}_{-0.029}$
& $-3.069\mkern4mu{}^{+0.031}_{-0.031}$
& $0.276$ \\

\addlinespace[0.2em]
\midrule
\addlinespace[0.2em]

M51
& B
& $1.071\mkern4mu{}^{+0.062}_{-0.074}$
& $-3.360\mkern4mu{}^{+0.127}_{-0.109}$
& $0.103$ \\

\bottomrule
\end{tabular}
\end{table*}
We then compute the signed residuals from the best-fit line, 
\(Res = \log\left(\Sigma_\mathrm{SFR, true}\right)-\log\left(\Sigma_\mathrm{SFR, pred}\right) = \log\left(\dfrac{\Sigma_\mathrm{SFR, true}}{\Sigma_\mathrm{SFR, pred}}\right)\), 
and examine their dependence on $\Delta v$. If a dependence exists, this means that $\Delta v$ is a hidden parameter in the KS relation.
\footnote{Note that since the method of fitting was ordinary least-squares, the linear covariance (and/or correlation) of the residuals of the fit ($Res$) with $\log\Sigma_{\rm mol}$ is zero. Correlation between $Res$ and non-linear functions of \(\log\Sigma_{\rm mol}\) is not expected, since that would require that the KS relation possess appreciable curvature in log-log space, while the KS relations analyzed in this work are well described by a linear relation in log-log space over the fitted ranges and show no appreciable curvature of the kind required for this effect.  
Hence, any dependence of $Res$ on $\Delta v$ cannot be explained by the correlation between $\Delta v$ and $\Sigma_{\rm mol}$.}

In group A, no significant correlation is found (see Fig.~\ref{fig:KS_DV_err_along}, right column).
In group B, the residuals show a systematic decrease with $\log\left(\Delta v\right)$, which we quantify by (least-squares) fitting a linear relation of the form $Res = a_0\log\left(\Delta v\right) + b_0$ (see Fig.~\ref{fig:KS_DV_err_perp}, right column). The uncertainties of the coefficients are determined similarly to the uncertainties of the coefficients of the KS relation, as described above. For consistency, we also perform the same fit on galaxies of group A. We assess the strength and significance of the correlation using the Spearman rank coefficient, $\rho$, and the coefficient of determination, $R^2$.

For all surveys, group B displays a weak but statistically significant dependence of $Res$ on $\Delta v$, with $R^2$ values in the range $0.05 \lesssim R^2 \lesssim 0.2$ and (negative) Spearman coefficients of $0.2 \lesssim |\rho| \lesssim 0.4$, all associated with $p<10^{-3}$. We note that the reported Spearman coefficients and $p$-values do not account for the non-independence of the pixels and, particularly, that the low $p$-values are likely due to the large number of data points. Hence, to assess the significance of the Spearman coefficients, we perform bootstraps similar to those used to assess the uncertainties of the KS and $Res-\Delta v$ fitted coefficients. This analysis confirms that $\rho$ is consistently negative, supporting the anticorrelation between $Res$ and $\Delta v$. This indicates that $\Delta v$ acts as a hidden parameter in the KS relation of group B.
Moreover, the fitted slopes, $a_0$, are similar across all surveys (the largest pairwise separation is $\sim 2\sigma$, between ALMaQUEST and EDGE--CALIFA), suggesting a common underlying trend, despite the differences in sample selection and observational methodology.
In contrast, group A values of $R^2$ and $\rho$ are both close to zero ($R^2 \lesssim 0.01$, $|\rho| \lesssim 0.06$), confirming the absence of any correlation. The fit parameters and correlation metrics for both groups in all surveys are summarized in Table~\ref{tab:all_fit_params}. 

At the spatial resolution of the surveys used in this work ($\sim$1 kpc), the measured CO velocity dispersion may contain contributions from unresolved rotation, beam smearing, shear, and other large-scale motions. These contributions may depend on the viewing angle of the galaxy. In Appendix~\ref{app:inclination}, we show that the observed trend of decreasing $Res$ with increasing $\Delta v$ does not depend on galaxy inclination. Additionally, the high-SFR values come largely from the centers of the galaxies, where extinction corrections become most uncertain. However, since group B behavior can be found throughout the range of $\Sigma_{\rm SFR}$ in the KS plots and since only a very small number of pixels belong to the centers, we believe that these uncertainties in the extinction correction cannot significantly affect our results.

Additionally, because the classification into group A and group B is itself based on the strength of local $Res - \Delta v$ trends, the possibility of a selection effect must be considered. Even if no true $Res - \Delta v$ dependence were present, random fluctuations would produce some locally negative fitted slopes. Since the classifier is designed to identify galaxies with the strongest and most persistent negative local slopes, it could preferentially select the tail of this noise distribution as group B. The same selected galaxies could then exhibit a negative pooled $Res - \Delta v$ slope simply as a consequence of this selection. To test whether this effect can account for the observed group B signal, we perform a permutation null test. For each galaxy, the measured \(\Delta v\) values are randomly permuted among all spatial measurements, while \(\Sigma_{\rm mol}\) and \(\Sigma_{\rm SFR}\) are left unchanged. For each of 2000 realizations, we rerun the complete classification procedure, including the calculation of \(S_3\), \(S_4\), and \(S_5\) and the GMM fit, and then recompute the pooled group B $Res - \Delta v$ slope \(a_0\). We consider a null realization to reproduce or exceed the observed signal if it produces both at least as many group B galaxies as observed (\(N_B\geq13\)) and a group B $Res-\Delta v$ slope at least as negative as the observed value. Only 6 of 2000 realizations satisfy both conditions in ALMaQUEST, and independently only 6 of 2000 do so in EDGE--CALIFA. Using the finite-permutation estimate \(p=(k+1)/(N_{\rm perm}+1)\), this corresponds to \(p=0.0035\) in each survey. We therefore conclude that the observed combination of the number of group B galaxies and the strength of their pooled $Res - \Delta v$ dependence is unlikely to arise from the selection procedure acting on random fluctuations alone.
\begin{table*}
\centering
\caption{Fitted parameters of the $Res - \Delta v$ relation of the two groups of galaxies }
\label{tab:all_fit_params}

\renewcommand{\arraystretch}{2.0}

\begin{tabular}{ccccccc}
\toprule
Survey
& Group
& $a_0$
& $b_0$
& $R^2$
& Spearman $\rho$
& $p$-value \\
\midrule

\multirow[c]{2}{*}[-0.4ex]{ALMaQUEST}
& A
& $-0.077\mkern4mu{}^{+0.049}_{-0.042}$
& $ 0.077\mkern4mu{}^{+0.041}_{-0.049}$
& $0.004$
& $-0.023\mkern4mu{}^{+0.036}_{-0.031}$
& $10^{-3}$ \\

& B
& $-0.576\mkern4mu{}^{+0.170}_{-0.114}$
& $ 0.650\mkern4mu{}^{+0.132}_{-0.189}$
& $0.187$
& $-0.444\mkern4mu{}^{+0.116}_{-0.066}$
& $<10^{-3}$ \\

\addlinespace[0.2em]
\midrule
\addlinespace[0.2em]

\multirow[c]{2}{*}[-0.4ex]{EDGE--CALIFA}
& A
& $ 0.006\mkern4mu{}^{+0.039}_{-0.038}$
& $-0.008\mkern4mu{}^{+0.052}_{-0.054}$
& $<10^{-3}$
& $-0.053\mkern4mu{}^{+0.037}_{-0.035}$
& $<10^{-3}$ \\

& B
& $-0.268\mkern4mu{}^{+0.065}_{-0.058}$
& $ 0.356\mkern4mu{}^{+0.077}_{-0.084}$
& $0.048$
& $-0.220\mkern4mu{}^{+0.050}_{-0.061}$
& $<10^{-3}$ \\

\addlinespace[0.2em]
\midrule
\addlinespace[0.2em]

PHANGS--ALMA
& A
& $ 0.132\mkern4mu{}^{+0.043}_{-0.045}$
& $-0.076\mkern4mu{}^{+0.026}_{-0.024}$
& $0.012$
& $ 0.121\mkern4mu{}^{+0.035}_{-0.039}$
& $<10^{-3}$ \\

\addlinespace[0.2em]
\midrule
\addlinespace[0.2em]

M51
& B
& $-0.360\mkern4mu{}^{+0.121}_{-0.109}$
& $ 0.353\mkern4mu{}^{+0.109}_{-0.120}$
& $0.078$
& $-0.229\mkern4mu{}^{+0.103}_{-0.093}$
& $<10^{-3}$ \\
\addlinespace[0.2em]
\bottomrule
\end{tabular}
\end{table*}

\section{Extending the Kennicutt-Schmidt Relation Including Velocity Dispersion}

For group B galaxies and for each survey separately, we perform a least-squares fit of the form 
\begin{equation}
    \log \Sigma_{\rm SFR} = n\log \Sigma_{\rm mol} + a\log \Delta v + C\,.
\end{equation}
We present the fitted parameters as well as the dispersion in $\log\Sigma_\mathrm{SFR}$ ($\sigma$) of the fit in Table~\ref{tab:all_3d_fit_params}. In all surveys, we consistently find $a$ to be negative, indicating (again) that, at fixed $\Sigma_\mathrm{mol}$, $\Sigma_\mathrm{SFR}$ decreases with increasing $\Delta v$.

The inclusion of $\Delta v$ results in only a marginal reduction of the overall scatter ($\sigma$) relative to the classical KS relation. 
This behavior is expected. 
If a fraction of the KS scatter $\sigma_\mathrm{KS}$ is explained by the dependence on $\Delta v$, with associated variance $\sigma_{\mathrm{KS}, \ \Delta v}$, the remaining scatter $\sigma_\mathrm{new}$ satisfies $\sigma_{\rm new}^2 = \sigma_{\rm KS}^2 - \sigma_{\rm KS, \ \Delta v}^2$.
Since the scatter adds in quadrature, a non-negligible fraction (e.g., 10\%) of the KS variance explained by $\Delta v$ may still lead to only a small reduction in the total scatter.

Nevertheless, the best-fit coefficients $n$ and $a$ are consistent across the different surveys, despite their independent observations, galaxy samples, and measurement methodologies. 
To quantify the survey-to-survey consistency of the best-fit coefficients, we compute the pairwise separations of the coefficients in units of their combined uncertainties. The coefficients $n$ of $\Sigma_{\rm mol}$ differ by $(0.3, 0.7, 0.4)\sigma$, corresponding to the pairs (ALMaQUEST---EDGE--CALIFA, ALMaQUEST---M51, EDGE--CALIFA---M51), and the coefficients $a$ of $\Delta v$ differ by $(1.7, 0.5, 0.9)\sigma$ (in the same survey order).

Given this close agreement between the coefficients of both $\Sigma_\mathrm{mol}$ and $\Delta v$ in the extended relations of the ALMaQUEST and EDGE--CALIFA surveys, we compute the equal-weight means of $n$ and $a$ of these two surveys (we do not include M51, which is a single galaxy). 
This yields $\bar n = 1.15^{+0.06}_{-0.07}$ and $\bar a = -0.59^{+0.08}_{-0.10}$.
The uncertainties were derived from the 16th and 84th percentiles of the distribution obtained by independently pairing bootstrap realizations from the two surveys and averaging their coefficient vectors. The corresponding scaling relation can therefore be written as
\begin{equation}\label{eq:3d_result}
    \Sigma_{\rm SFR} \propto \Sigma_{\rm mol}^{\, 1.15} \, \Delta v^{-0.59}.
\end{equation}

Surprisingly, \citet{Wang2020}, who fit a relation of the same form [see their Eq. (9)] to the combined sample of the six galaxies they use, find a very similar exponent for $\Delta v$ ($a=-0.61\pm0.11$), as the one in our Eq. \eqref{eq:3d_result} (although their CO-luminosity exponent, $n=0.84\pm0.06$, is shallower than our molecular-gas exponent). However, they did not apply our group A/group B classification, hence, this close agreement applies to their full sample fit.

The consistently negative \(\Delta v\) coefficients obtained independently in ALMaQUEST and EDGE–CALIFA provide a strong indication that \(\Delta v\) plays a role in regulating star formation at fixed \(\Sigma_{\rm mol}\) in certain galaxies. The similarity of the fitted coefficients between the two surveys further supports this conclusion, since the coefficients encode the underlying physics of the process of star formation.

\renewcommand{\arraystretch}{2.0}
\begin{table*}
\centering
\caption{Fitted parameters of the extended relation for group B}
\label{tab:all_3d_fit_params}
\begin{tabular}{ccccc}
\toprule
     Survey &                        $C$ &                       $n$ &                        $a$ & $\sigma$ \\
\midrule
  ALMaQUEST & $-2.133^{+0.140}_{-0.127}$ & $1.128^{+0.105}_{-0.090}$ & $-0.741^{+0.128}_{-0.152}$ &  $0.347$ \\
  \midrule
EDGE-CALIFA & $-2.760^{+0.234}_{-0.179}$ & $1.164^{+0.073}_{-0.100}$ & $-0.435^{+0.104}_{-0.118}$ &  $0.338$ \\
\midrule
        M51 & $-2.987^{+0.152}_{-0.141}$ & $1.209^{+0.063}_{-0.082}$ & $-0.636^{+0.190}_{-0.168}$ &  $0.096$ \\
\bottomrule
\end{tabular}
\end{table*}

\section{Comparison of the Global and Resolved Properties of the Two Groups}

\begin{figure*}
  \centering
  \includegraphics[width=\hsize]{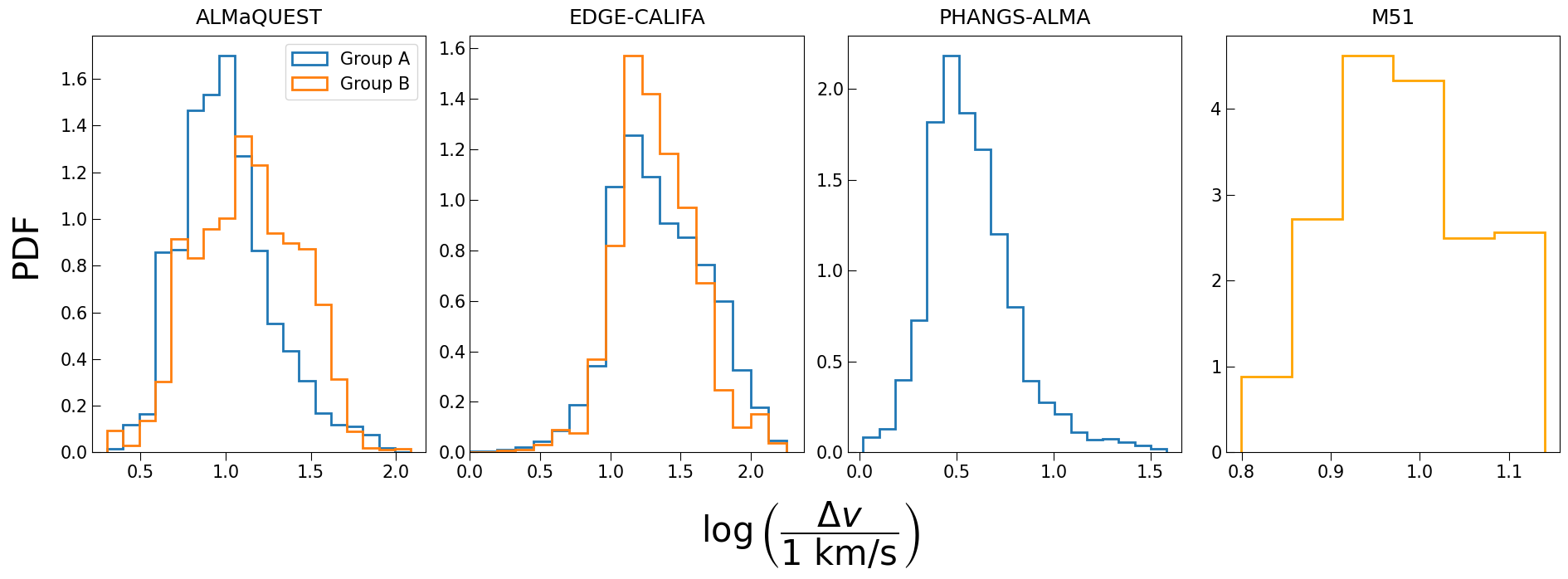}
  \caption{PDFs of $\log \Delta v$ for the two groups across all surveys. No systematic differences are identified between the two groups.}
  \label{fig:Dv_PDFs}
\end{figure*}

\begin{figure*}
  \centering
  \includegraphics[width=\hsize]{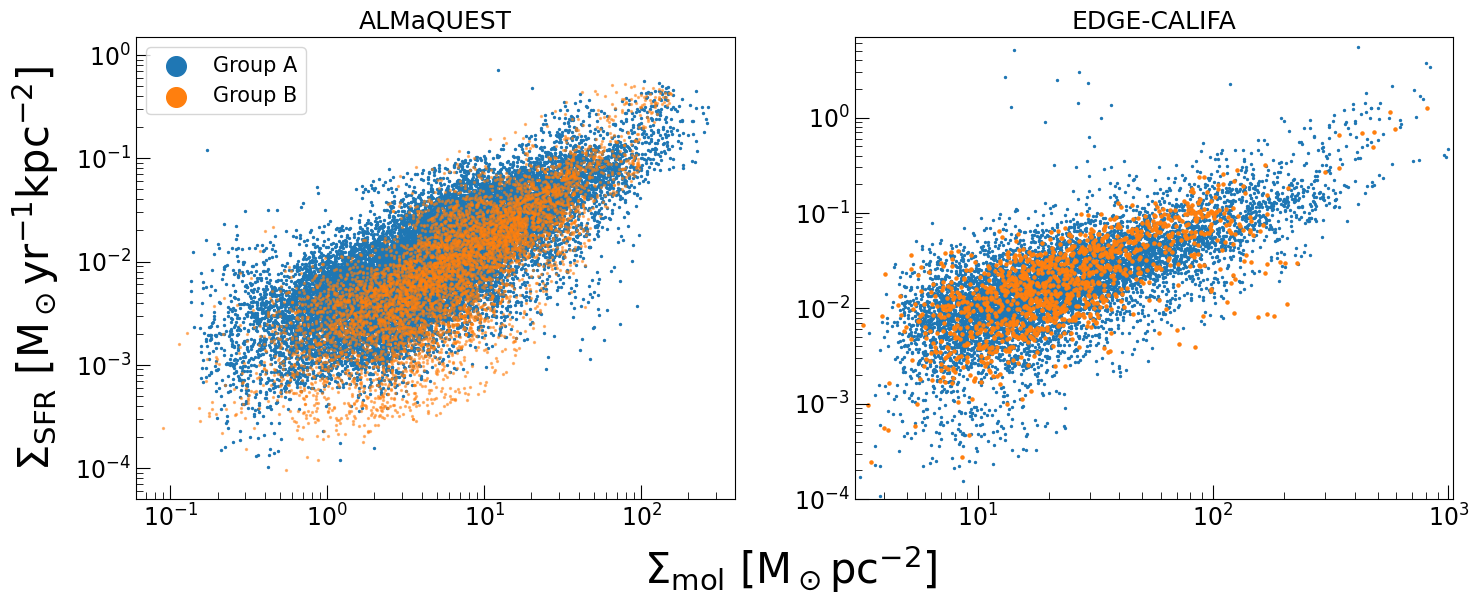}
  \caption{KS relation for group A (blue points) and group B (orange points) in the ALMaQUEST and EDGE-CALIFA surveys. Group B does not seem to exhibit different behavior than group A.}
  \label{fig:KS_law_2groups}
\end{figure*}

\begin{figure*}
  \centering
  \includegraphics[width=\hsize]{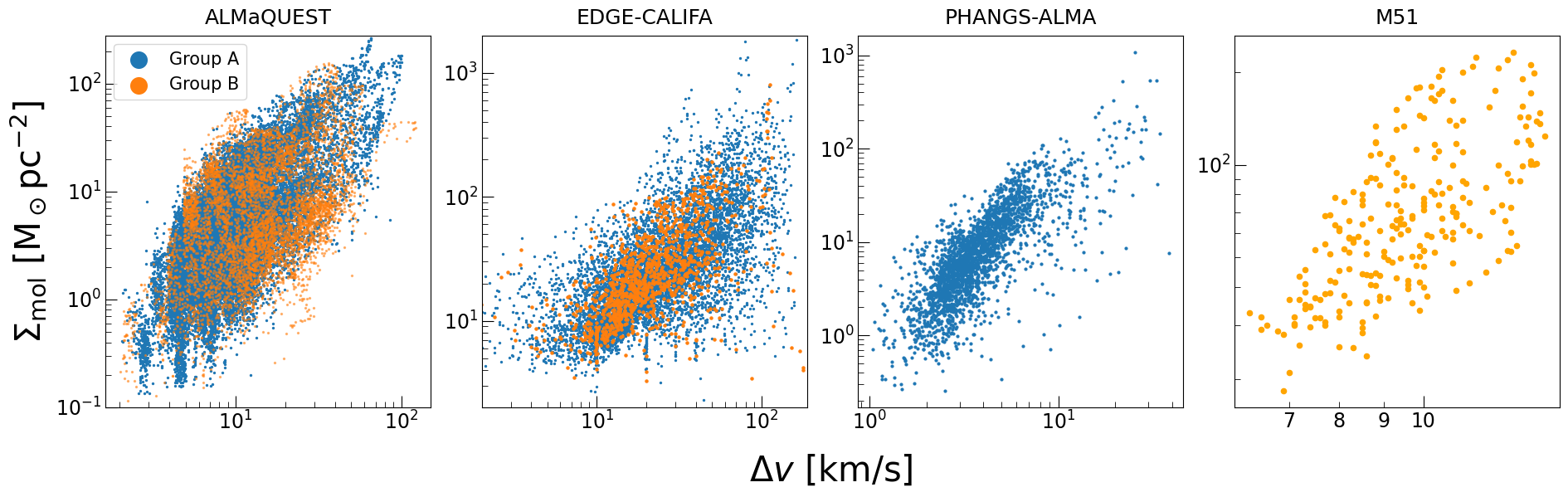}
  \caption{$\Sigma_\mathrm{mol}$--$\Delta v$ relation for group A (blue points) and group B (orange points) in all surveys. Group B does not seem to exhibit different behavior than group A.}
  \label{fig:Smol_Dv}
\end{figure*}

\begin{figure*}
  \centering
  \includegraphics[width=\hsize]{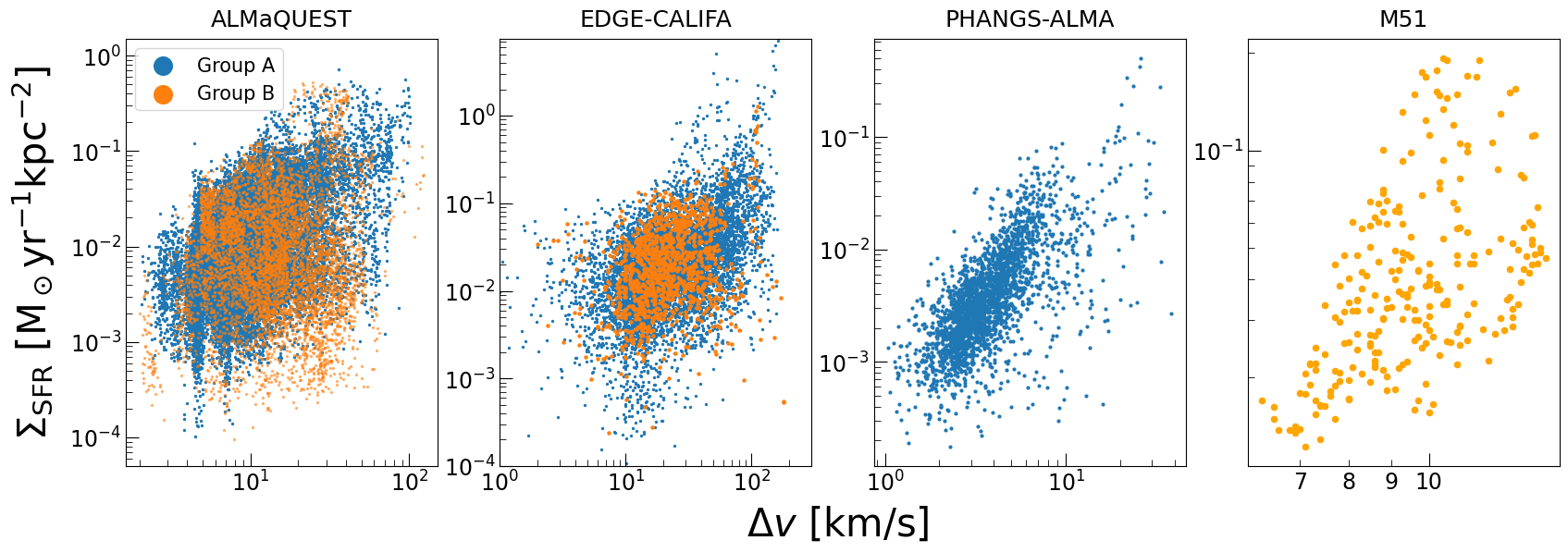}
  \caption{$\Sigma_\mathrm{SFR}$--$\Delta v$ relation for group A (blue points) and group B (orange points) in all surveys. In ALMaQUEST and EDGE-CALIFA, for $\Delta v$ higher than $\sim 20$ km/s, group B galaxies have slightly lower $\Sigma_\mathrm{SFR}$ than group A galaxies.}
  \label{fig:sSFR_Dv}
\end{figure*}

Having established a distinct behavior of the two groups regarding the role of $\Delta v$ in the KS relation, we now investigate whether this difference in behavior is reflected in their properties. We perform a series of comparative tests between the two groups, examining global and spatially resolved quantities to assess whether they differ systematically. As we show below, we do not find evidence for a clear distinction between the two groups based on these properties.

\subsection{Comparison of \texorpdfstring{$\Delta v$}{velocity-dispersion} distributions}
We compare the distributions of $\Delta v$ for the two groups by constructing their corresponding probability density functions (PDFs) (see Fig.~\ref{fig:Dv_PDFs}). We find that the distributions are similar in shape. In ALMaQUEST, the PDF of group A is more localized than the PDF of group B, while in EDGE--CALIFA, this is reversed. Thus, we cannot point to any systematic difference in the distributions of $\Delta v$ of the two groups.

\subsection{Comparison of KS, \texorpdfstring{$\Sigma_{\mathrm{mol}}$--$\Delta v$}{molecular-gas--velocity-dispersion}, and \texorpdfstring{$\Sigma_{\mathrm{SFR}}$--$\Delta v$}{SFR--velocity-dispersion} relations}

We compare the KS, $\Sigma_{\mathrm{mol}}$–$\Delta v$, and $\Sigma_{\mathrm{SFR}}$–$\Delta v$ relations of the two groups.
The KS and $\Sigma_{\mathrm{mol}}$–$\Delta v$ relations are consistent between groups A and B across all surveys (see Figs.~\ref{fig:KS_law_2groups},~\ref{fig:Smol_Dv}). In contrast, the $\Sigma_{\mathrm{SFR}}$–$\Delta v$ relation shows slight differences between the two groups in the ALMaQUEST and EDGE--CALIFA surveys: for $\Delta v$ higher than $\sim 20$ km/s, group B galaxies have slightly lower $\Sigma_\mathrm{SFR}$ than group A galaxies (see Fig.~\ref{fig:sSFR_Dv}), indicating suppression of star formation in regions with higher $\Delta v$. For completeness, we also include in Fig.~\ref{fig:sSFR_Dv} the PHANGS--ALMA galaxies and M51. Note that in M51, values of $\Delta v$ are much lower, since they do not include bulk motions.

\subsection{AGN Content}
We compared the AGN fractions of the two groups across the available surveys. For the ALMaQUEST sample, the MaNGA AGN Catalog \citep{Comerford2024}, which is based on spatially resolved emission-line diagnostics with conservative criteria for nuclear activity, yields no AGN in either group. According to an earlier version of this catalog \citep{Comerford2020}, which used a more liberal class of radio sources that could include possible contamination by star formation rather than AGN, there are three AGN in each group, corresponding to 23\% in group B and 9\% in group A. The \citet{Alban2023} catalog, which identifies AGN via emission-line diagnostics with additional constraints on ionization sources, yields one object in group B and none in group A.

For the EDGE-CALIFA sample, we used the \citet{Lacerda2020} catalog, in which AGN are identified using optical emission-line (BPT) classifications, typically separating Seyfert and LINER-like nuclei from star-forming regions. We find six AGN in group A and none in group B.

We do not perform this analysis for the PHANGS-ALMA sample, as no uniform AGN catalog exists.
In addition, the PHANGS-ALMA sample was explicitly selected to target normal, star-forming disk galaxies and to avoid systems with strong or disruptive AGN activity \citep{Leroy2021}.

Overall, we find no robust evidence for a systematic difference in AGN content between the two groups of galaxies.

\subsection{Mergers}
The PHANGS--ALMA sample does not include mergers by design \citep{Leroy2021}.
In ALMaQUEST, mergers are twice as frequent in group B as in group A (4 galaxies, corresponding to 30\% in group B vs. 5 galaxies, corresponding to 15\% in group A), and M51 (which is in group B) is known to be an interacting galaxy. This indicates that interactions may affect the properties of the interstellar medium and star formation, giving rise to the dependence of $\Sigma_\mathrm{SFR}$ on $\Delta v$. 
Interestingly, a recent spatially resolved study of luminous infrared galaxies along the merger sequence \citep{SanchezGarcia2026} has found that, in intermediate stages of interactions, the star formation efficiency decreases with increasing velocity dispersion within molecular clumps, which is consistent with our observed trend.
However, since the $\Delta v$ distributions of groups A and B are similar, this difference in merger fraction does not reflect systematically higher velocity dispersions as a result of interactions in group B. 
Given the small number of merger systems, this trend should be interpreted with caution.

We do not perform the same analysis for the EDGE--CALIFA survey, since no uniform catalog of mergers exists for this particular sample of galaxies.

\subsection{Starbursts}
The PHANGS--ALMA sample does not include starburst galaxies \citep{Leroy2021} and M51 is not a starburst.
In the ALMaQUEST survey, starburst galaxies are twice as rare in group B (2 galaxies, corresponding to 15\% in group B vs. 10 galaxies, corresponding to 30\% in group A).
This indicates that the observed trend of decreasing $\Sigma_\mathrm{SFR}$ with increasing $\Delta v$ in group B galaxies is unlikely due to star formation feedback. Again, we emphasize that because of the small-number statistics, this is a hint rather than a definitive statement.

We do not perform the same analysis for the EDGE-CALIFA survey, since no uniform catalog of starbursts exists for this particular sample of galaxies.
\section{Cloud–Cloud Motions as a Source of Star Formation Suppression}

Since $\Delta v$ measurements in the ALMaQUEST and EDGE-CALIFA surveys are at a spatial scale of $\sim$1 kpc, each region typically contains many clouds. Thus, the measured velocity dispersion is unlikely to trace internal turbulence of individual clouds, which is usually a few km/s, especially since we observe $\Delta v$ values often exceeding 100 km/s. It is therefore plausible that $\Delta v$ in these datasets predominantly reflects relative motions between clouds (``cloud–cloud motions'') rather than internal turbulence. Under this interpretation, the observed result---that for fixed $\Sigma_\mathrm{mol}$, regions with higher $\Delta v$ have lower $\Sigma_\mathrm{SFR}$---is consistent with cloud--cloud motions suppressing or hindering star formation efficiency. This result is unexpected, since most of the simulations in the literature support the opposite, i.e., that cloud--cloud collisions enhance star formation (e.g., \citealp{Fukui2021}).
Some exceptions exist, however. For example, in strongly barred galaxies, \citet{Fujimoto2020} found that fast cloud--cloud collisions in the bar region correlate with a suppression of massive star formation. 

However, since there is no systematic difference in the $\Delta v$ distributions of the two groups, it is counterintuitive that this explanation of cloud--cloud motions hindering star formation applies to group B galaxies only.
The absence of a correlation between $\Delta v$ and $\Sigma_{\mathrm{SFR}}$ in group A suggests that velocity dispersion itself might not be the fundamental driver of the observed suppression of star formation. Instead, $\Delta v$ may act as a proxy for more specific properties of the gas velocity field, such as shear, dispersive motions, or other large-scale kinematic structures. Also, $\Delta v$ may act as a proxy for other physical mechanisms, such as magnetic fields [through Alfv\'{e}nic turbulence \citep{MTK06}], that support clouds against gravitational collapse. In this framework, the distinction between the two groups may reflect differences in the dominant type of gas motions: in one regime, $\Delta v$ traces kinematic components that suppress star formation, while in the other it is dominated by motions that do not significantly affect star formation efficiency. A more complete characterization of the velocity field is required to fully capture its role in regulating star formation.

\section{Summary and Conclusions}

We examined the role of molecular gas velocity dispersion ($\Delta v$) in shaping the resolved Kennicutt-Schmidt (KS) relation using data from ALMaQUEST, EDGE--CALIFA, PHANGS--ALMA, and M51. Galaxies were divided into two groups based on the orientation of their $\Delta v$ gradients relative to the KS line: group A, with gradients aligned with the KS line, and group B, with gradients having a component perpendicular to it.

Group A shows no significant correlation between KS residuals and $\Delta v$, whereas group B exhibits a weak but statistically significant linear dependence of the residuals on $\log \Delta v$, with residuals decreasing systematically with increasing $\log \Delta v$. The correlation strength and slope are consistent across all surveys (Spearman $|\rho|\approx 0.2 - 0.4$, $p<10^{-3}$, $R^2\sim0.1$, $|a_0|\approx 0.3 - 0.6$), suggesting a common physical mechanism. At fixed $\Sigma_\mathrm{mol}$, regions with higher $\Delta v$ have systematically lower $\Sigma_\mathrm{SFR}$, implying that $\Delta v$ acts as a secondary parameter regulating star formation efficiency.

For group B galaxies, we extended the KS relation, fitting a relation of the form $\Sigma_\mathrm{SFR} \propto \Sigma_\mathrm{mol}^n \  \Delta v^{a}$ to the data of each survey separately. Across both the ALMaQUEST and EDGE--CALIFA surveys, we consistently find a negative exponent $a$, indicating that $\Sigma_\mathrm{SFR}$ decreases with increasing $\Delta v$ at fixed $\Sigma_\mathrm{mol}$. 
 Notably, the fitted coefficients $n$ and $a$ are within $0.3\sigma$ and $1.7\sigma$ of each other, respectively, despite systematic differences between the two surveys.
The consistently negative \(\Delta v\) coefficients obtained independently in ALMaQUEST and EDGE–CALIFA provide a strong indication that \(\Delta v\) plays a role in regulating star formation. Since the exponents encode the underlying physics governing star formation, the similarity of the fitted coefficients between the two surveys provides an additional consistency check.

Despite this coherent trend, we find no single property--such as $\Delta v$ distribution or AGN content--that clearly differentiates group B from group A. 
Mergers are found to be approximately twice as frequent in group B compared to group A in the ALMaQUEST sample. However, since the $\Delta v$ distributions are similar between the two groups, this difference does not reflect systematically higher velocity dispersions and may instead point to a more subtle link between interactions and the role of $\Delta v$ in the KS relation. Given the small number of mergers, this result should be interpreted with caution.
Overall, the physical origin of this difference in the behavior of the two groups remains uncertain and warrants further investigation with higher-resolution kinematic data, as well as more detailed knowledge of the velocity field of the galaxies.

 \begin{acknowledgements}
     We would like to thank Li-Hwai Lin for sharing the ALMaQUEST Data Products with us. We would also like to thank Bolatto et al.\ for making the EDGE-CALIFA data available and Leroy et al.\ for making the PHANGS-ALMA as well as the M51 data available.
     We thank an anonymous referee for insightful comments that significantly improved the manuscript.
     This research is partially funded by the ERC grant, mw-atlas project no. 101166905. Views and opinions expressed are, however, those of the author(s) only and do not necessarily reflect those of the European Union or the European Research Council Executive Agency. Neither the European Union nor the granting authority can be held responsible for them.
 \end{acknowledgements}

\bibliographystyle{aa}
\bibliography{references}

@ARTICLE{Schmidt1959,
       author = {{Schmidt}, Maarten},
        title = "{The Rate of Star Formation.}",
      journal = {\apj},
         year = 1959,
        month = mar,
       volume = {129},
        pages = {243},
          doi = {10.1086/146614},
       adsurl = {https://ui.adsabs.harvard.edu/abs/1959ApJ...129..243S}
}

@ARTICLE{Kennicutt1989,
       author = {{Kennicutt}, Jr., Robert C.},
        title = "{The Star Formation Law in Galactic Disks}",
      journal = {\apj},
         year = 1989,
        month = sep,
       volume = {344},
        pages = {685},
          doi = {10.1086/167834},
       adsurl = {https://ui.adsabs.harvard.edu/abs/1989ApJ...344..685K}
}

@ARTICLE{Kennicutt1998,
       author = {{Kennicutt}, Jr., Robert C.},
        title = "{The Global Schmidt Law in Star-forming Galaxies}",
      journal = {\apj},
         year = 1998,
        month = may,
       volume = {498},
       number = {2},
        pages = {541-552},
          doi = {10.1086/305588},
archivePrefix = {arXiv},
       eprint = {astro-ph/9712213},
 primaryClass = {astro-ph},
       adsurl = {https://ui.adsabs.harvard.edu/abs/1998ApJ...498..541K}
}

@ARTICLE{Bigiel2008,
       author = {{Bigiel}, F. and {Leroy}, A. and {Walter}, F. and {Brinks}, E. and {de Blok}, W.~J.~G. and {Madore}, B. and {Thornley}, M.~D.},
        title = "{The Star Formation Law in Nearby Galaxies on Sub-Kpc Scales}",
      journal = {\aj},
         year = 2008,
        month = dec,
       volume = {136},
       number = {6},
        pages = {2846-2871},
          doi = {10.1088/0004-6256/136/6/2846},
archivePrefix = {arXiv},
       eprint = {0810.2541},
 primaryClass = {astro-ph},
       adsurl = {https://ui.adsabs.harvard.edu/abs/2008AJ....136.2846B}
}

@article{Leroy2008,
  author = {{Leroy}, Adam K. and {Walter}, Fabian and {Brinks}, Elias and {Bigiel}, Frank and {de Blok}, W.~J.~G. and {Madore}, Barry and {Thornley}, M.~D.},
        title = "{The Star Formation Efficiency in Nearby Galaxies: Measuring Where Gas Forms Stars Effectively}",
      journal = {\aj},
         year = 2008,
        month = dec,
       volume = {136},
       number = {6},
        pages = {2782-2845},
          doi = {10.1088/0004-6256/136/6/2782},
archivePrefix = {arXiv},
       eprint = {0810.2556},
 primaryClass = {astro-ph},
       adsurl = {https://ui.adsabs.harvard.edu/abs/2008AJ....136.2782L}
}

@article{Schruba2011,
  author = {{Schruba}, Andreas and {Leroy}, Adam K. and {Walter}, Fabian and {Bigiel}, Frank and {Brinks}, Elias and {de Blok}, W.~J.~G. and {Dumas}, Gaelle and {Kramer}, Carsten and {Rosolowsky}, Erik and {Sandstrom}, Karin and {Schuster}, Karl and {Usero}, Antonio and {Weiss}, Axel and {Wiesemeyer}, Helmut},
        title = "{A Molecular Star Formation Law in the Atomic-gas-dominated Regime in Nearby Galaxies}",
      journal = {\aj},
         year = 2011,
        month = aug,
       volume = {142},
       number = {2},
          eid = {37},
        pages = {37},
          doi = {10.1088/0004-6256/142/2/37},
archivePrefix = {arXiv},
       eprint = {1105.4605},
 primaryClass = {astro-ph.CO},
       adsurl = {https://ui.adsabs.harvard.edu/abs/2011AJ....142...37S}
}

@article{Silk1997,
  author = {{Silk}, Joseph},
        title = "{Feedback, Disk Self-Regulation, and Galaxy Formation}",
      journal = {\apj},
         year = 1997,
        month = may,
       volume = {481},
       number = {2},
        pages = {703-709},
          doi = {10.1086/304073},
archivePrefix = {arXiv},
       eprint = {astro-ph/9612117},
 primaryClass = {astro-ph},
       adsurl = {https://ui.adsabs.harvard.edu/abs/1997ApJ...481..703S}
}

@INPROCEEDINGS{Elmegreen1997,
       author = {{Elmegreen}, B.~G.},
        title = "{Theory of Starbursts in Nuclear Rings}",
    booktitle = {Revista Mexicana de Astronomia y Astrofisica Conference Series},
         year = 1997,
       editor = {{Franco}, J. and {Terlevich}, R. and {Serrano}, A.},
       series = {Revista Mexicana de Astronomia y Astrofisica Conference Series},
       volume = {6},
        month = may,
        pages = {165},
       adsurl = {https://ui.adsabs.harvard.edu/abs/1997RMxAC...6..165E}
}

@article{Tan2000,
  author = {{Tan}, Jonathan C.},
        title = "{Star Formation Rates in Disk Galaxies and Circumnuclear Starbursts from Cloud Collisions}",
      journal = {\apj},
         year = 2000,
        month = jun,
       volume = {536},
       number = {1},
        pages = {173-184},
          doi = {10.1086/308905},
archivePrefix = {arXiv},
       eprint = {astro-ph/9906355},
 primaryClass = {astro-ph},
       adsurl = {https://ui.adsabs.harvard.edu/abs/2000ApJ...536..173T}
}

@article{Ostriker2010,
  author = {{Ostriker}, Eve C. and {McKee}, Christopher F. and {Leroy}, Adam K.},
        title = "{Regulation of Star Formation Rates in Multiphase Galactic Disks: A Thermal/Dynamical Equilibrium Model}",
      journal = {\apj},
         year = 2010,
        month = oct,
       volume = {721},
       number = {2},
        pages = {975-994},
          doi = {10.1088/0004-637X/721/2/975},
archivePrefix = {arXiv},
       eprint = {1008.0410},
 primaryClass = {astro-ph.CO},
       adsurl = {https://ui.adsabs.harvard.edu/abs/2010ApJ...721..975O}
}

@article{Tassis2007,
   author = {{Tassis}, Konstantinos},
        title = "{The star formation law in a multifractal ISM}",
      journal = {\mnras},
         year = 2007,
        month = dec,
       volume = {382},
       number = {3},
        pages = {1317-1323},
          doi = {10.1111/j.1365-2966.2007.12472.x},
archivePrefix = {arXiv},
       eprint = {0709.1474},
 primaryClass = {astro-ph},
       adsurl = {https://ui.adsabs.harvard.edu/abs/2007MNRAS.382.1317T}
}

@article{Lin2020,
  author = {{Lin}, Lihwai and {Ellison}, Sara L. and {Pan}, Hsi-An and {Thorp}, Mallory D. and {Su}, Yung-Chau and {S{\'a}nchez}, Sebasti{\'a}n F. and {Belfiore}, Francesco and {Bothwell}, M.~S. and {Bundy}, Kevin and {Chen}, Yan-Mei and {Concas}, Alice and {Hsieh}, Bau-Ching and {Hsieh}, Pei-Ying and {Li}, Cheng and {Maiolino}, Roberto and {Masters}, Karen and {Newman}, Jeffrey A. and {Rowlands}, Kate and {Shi}, Yong and {Smethurst}, Rebecca and {Stark}, David V. and {Xiao}, Ting and {Yu}, Po-Chieh},
        title = "{ALMaQUEST. IV. The ALMA-MaNGA QUEnching and STar Formation (ALMaQUEST) Survey}",
      journal = {\apj},
         year = 2020,
        month = nov,
       volume = {903},
       number = {2},
          eid = {145},
        pages = {145},
          doi = {10.3847/1538-4357/abba3a},
archivePrefix = {arXiv},
       eprint = {2010.01751},
 primaryClass = {astro-ph.GA},
       adsurl = {https://ui.adsabs.harvard.edu/abs/2020ApJ...903..145L}
}

@article{Sun2022,
author = {{Sun}, Jiayi and {Leroy}, Adam K. and {Rosolowsky}, Erik and {Hughes}, Annie and {Schinnerer}, Eva and {Schruba}, Andreas and {Koch}, Eric W. and {Blanc}, Guillermo A. and {Chiang}, I-Da and {Groves}, Brent and {Liu}, Daizhong and {Meidt}, Sharon and {Pan}, Hsi-An and {Pety}, J{\'e}r{\^o}me and {Querejeta}, Miguel and {Saito}, Toshiki and {Sandstrom}, Karin and {Sardone}, Amy and {Usero}, Antonio and {Utomo}, Dyas and {Williams}, Thomas G. and {Barnes}, Ashley T. and {Benincasa}, Samantha M. and {Bigiel}, Frank and {Bolatto}, Alberto D. and {Boquien}, M{\'e}d{\'e}ric and {Chevance}, M{\'e}lanie and {Dale}, Daniel A. and {Deger}, Sinan and {Emsellem}, Eric and {Glover}, Simon C.~O. and {Grasha}, Kathryn and {Henshaw}, Jonathan D. and {Klessen}, Ralf S. and {Kreckel}, Kathryn and {Kruijssen}, J.~M. Diederik and {Ostriker}, Eve C. and {Thilker}, David A.},
        title = "{Molecular Cloud Populations in the Context of Their Host Galaxy Environments: A Multiwavelength Perspective}",
      journal = {\aj},
         year = 2022,
        month = aug,
       volume = {164},
       number = {2},
          eid = {43},
        pages = {43},
          doi = {10.3847/1538-3881/ac74bd},
archivePrefix = {arXiv},
       eprint = {2206.07055},
 primaryClass = {astro-ph.GA},
       adsurl = {https://ui.adsabs.harvard.edu/abs/2022AJ....164...43S}
}

@article{Leroy2017,
  author = {{Leroy}, Adam K. and {Schinnerer}, Eva and {Hughes}, Annie and {Kruijssen}, J.~M. Diederik and {Meidt}, Sharon and {Schruba}, Andreas and {Sun}, Jiayi and {Bigiel}, Frank and {Aniano}, Gonzalo and {Blanc}, Guillermo A. and {Bolatto}, Alberto and {Chevance}, M{\'e}lanie and {Colombo}, Dario and {Gallagher}, Molly and {Garcia-Burillo}, Santiago and {Kramer}, Carsten and {Querejeta}, Miguel and {Pety}, Jerome and {Thompson}, Todd A. and {Usero}, Antonio},
        title = "{Cloud-scale ISM Structure and Star Formation in M51}",
      journal = {\apj},
         year = 2017,
        month = sep,
       volume = {846},
       number = {1},
          eid = {71},
        pages = {71},
          doi = {10.3847/1538-4357/aa7fef},
archivePrefix = {arXiv},
       eprint = {1706.08540},
 primaryClass = {astro-ph.GA},
       adsurl = {https://ui.adsabs.harvard.edu/abs/2017ApJ...846...71L}
}

@article{Leroy2021,
  author = {{Leroy}, Adam K. and {Schinnerer}, Eva and {Hughes}, Annie and {Rosolowsky}, Erik and {Pety}, J{\'e}r{\^o}me and {Schruba}, Andreas and {Usero}, Antonio and {Blanc}, Guillermo A. and {Chevance}, M{\'e}lanie and {Emsellem}, Eric and {Faesi}, Christopher M. and {Herrera}, Cinthya N. and {Liu}, Daizhong and {Meidt}, Sharon E. and {Querejeta}, Miguel and {Saito}, Toshiki and {Sandstrom}, Karin M. and {Sun}, Jiayi and {Williams}, Thomas G. and {Anand}, Gagandeep S. and {Barnes}, Ashley T. and {Behrens}, Erica A. and {Belfiore}, Francesco and {Benincasa}, Samantha M. and {Be{\v{s}}li{\'c}}, Ivana and {Bigiel}, Frank and {Bolatto}, Alberto D. and {den Brok}, Jakob S. and {Cao}, Yixian and {Chandar}, Rupali and {Chastenet}, J{\'e}r{\'e}my and {Chiang}, I-Da and {Congiu}, Enrico and {Dale}, Daniel A. and {Deger}, Sinan and {Eibensteiner}, Cosima and {Egorov}, Oleg V. and {Garc{\'\i}a-Rodr{\'\i}guez}, Axel and {Glover}, Simon C.~O. and {Grasha}, Kathryn and {Henshaw}, Jonathan D. and {Ho}, I. -Ting and {Kepley}, Amanda A. and {Kim}, Jaeyeon and {Klessen}, Ralf S. and {Kreckel}, Kathryn and {Koch}, Eric W. and {Kruijssen}, J.~M. Diederik and {Larson}, Kirsten L. and {Lee}, Janice C. and {Lopez}, Laura A. and {Machado}, Josh and {Mayker}, Ness and {McElroy}, Rebecca and {Murphy}, Eric J. and {Ostriker}, Eve C. and {Pan}, Hsi-An and {Pessa}, Ismael and {Puschnig}, Johannes and {Razza}, Alessandro and {S{\'a}nchez-Bl{\'a}zquez}, Patricia and {Santoro}, Francesco and {Sardone}, Amy and {Scheuermann}, Fabian and {Sliwa}, Kazimierz and {Sormani}, Mattia C. and {Stuber}, Sophia K. and {Thilker}, David A. and {Turner}, Jordan A. and {Utomo}, Dyas and {Watkins}, Elizabeth J. and {Whitmore}, Bradley},
        title = "{PHANGS-ALMA: Arcsecond CO(2-1) Imaging of Nearby Star-forming Galaxies}",
      journal = {\apjs},
         year = 2021,
        month = dec,
       volume = {257},
       number = {2},
          eid = {43},
        pages = {43},
          doi = {10.3847/1538-4365/ac17f3},
archivePrefix = {arXiv},
       eprint = {2104.07739},
 primaryClass = {astro-ph.GA},
       adsurl = {https://ui.adsabs.harvard.edu/abs/2021ApJS..257...43L}
}

@article{Bolatto2017,
  author = {{Bolatto}, Alberto D. and {Wong}, Tony and {Utomo}, Dyas and {Blitz}, Leo and {Vogel}, Stuart N. and {S{\'a}nchez}, Sebasti{\'a}n F. and {Barrera-Ballesteros}, Jorge and {Cao}, Yixian and {Colombo}, Dario and {Dannerbauer}, Helmut and {Garc{\'\i}a-Benito}, Rub{\'e}n and {Herrera-Camus}, Rodrigo and {Husemann}, Bernd and {Kalinova}, Veselina and {Leroy}, Adam K. and {Leung}, Gigi and {Levy}, Rebecca C. and {Mast}, Dami{\'a}n and {Ostriker}, Eve and {Rosolowsky}, Erik and {Sandstrom}, Karin M. and {Teuben}, Peter and {van de Ven}, Glenn and {Walter}, Fabian},
        title = "{The EDGE-CALIFA Survey: Interferometric Observations of 126 Galaxies with CARMA}",
      journal = {\apj},
         year = 2017,
        month = sep,
       volume = {846},
       number = {2},
          eid = {159},
        pages = {159},
          doi = {10.3847/1538-4357/aa86aa},
archivePrefix = {arXiv},
       eprint = {1704.02504},
 primaryClass = {astro-ph.GA},
       adsurl = {https://ui.adsabs.harvard.edu/abs/2017ApJ...846..159B}
}

@article{Fujimoto2020,
  author={Fujimoto, Yusuke and Maeda, Fumiya and Habe, Asao and Ohta, Kouji},
  journal={Monthly Notices of the Royal Astronomical Society}, 
  title={Fast cloud–cloud collisions in a strongly barred galaxy: suppression of massive star formation}, 
  year={2020},
  volume={494},
  number={1},
  pages={2131-2146},
  doi={10.1093/mnras/staa840}}

@article{Tully1977,
  author = {{Tully}, R.~B. and {Fisher}, J.~R.},
        title = "{A new method of determining distances to galaxies.}",
      journal = {\aap},
         year = 1977,
        month = feb,
       volume = {54},
        pages = {661-673},
       adsurl = {https://ui.adsabs.harvard.edu/abs/1977A&A....54..661T}
}

@article{Faber1976,
   author = {{Faber}, S.~M. and {Jackson}, R.~E.},
        title = "{Velocity dispersions and mass-to-light ratios for elliptical galaxies.}",
      journal = {\apj},
         year = 1976,
        month = mar,
       volume = {204},
        pages = {668-683},
          doi = {10.1086/154215},
       adsurl = {https://ui.adsabs.harvard.edu/abs/1976ApJ...204..668F}
}

@article{Ferrarese2000,
  author = {{Ferrarese}, Laura and {Merritt}, David},
        title = "{A Fundamental Relation between Supermassive Black Holes and Their Host Galaxies}",
      journal = {\apjl},
         year = 2000,
        month = aug,
       volume = {539},
       number = {1},
        pages = {L9-L12},
          doi = {10.1086/312838},
archivePrefix = {arXiv},
       eprint = {astro-ph/0006053},
 primaryClass = {astro-ph},
       adsurl = {https://ui.adsabs.harvard.edu/abs/2000ApJ...539L...9F}
}

@article{Djorgovski1987,
  author = {{Djorgovski}, S. and {Davis}, Marc},
        title = "{Fundamental Properties of Elliptical Galaxies}",
      journal = {\apj},
         year = 1987,
        month = feb,
       volume = {313},
        pages = {59},
          doi = {10.1086/164948},
       adsurl = {https://ui.adsabs.harvard.edu/abs/1987ApJ...313...59D}
}

@article{PAWS,
  author = {{Colombo}, Dario and {Hughes}, Annie and {Schinnerer}, Eva and {Meidt}, Sharon E. and {Leroy}, Adam K. and {Pety}, J{\'e}r{\^o}me and {Dobbs}, Clare L. and {Garc{\'\i}a-Burillo}, Santiago and {Dumas}, Ga{\"e}lle and {Thompson}, Todd A. and {Schuster}, Karl F. and {Kramer}, Carsten},
        title = "{The PdBI Arcsecond Whirlpool Survey (PAWS): Environmental Dependence of Giant Molecular Cloud Properties in M51}",
      journal = {\apj},
         year = 2014,
        month = mar,
       volume = {784},
       number = {1},
          eid = {3},
        pages = {3},
          doi = {10.1088/0004-637X/784/1/3},
archivePrefix = {arXiv},
       eprint = {1401.1505},
 primaryClass = {astro-ph.GA},
       adsurl = {https://ui.adsabs.harvard.edu/abs/2014ApJ...784....3C}
}

@article{manga1,
  author = {{Bundy}, Kevin and {Bershady}, Matthew A. and {Law}, David R. and {Yan}, Renbin and {Drory}, Niv and {MacDonald}, Nicholas and {Wake}, David A. and {Cherinka}, Brian and {S{\'a}nchez-Gallego}, Jos{\'e} R. and {Weijmans}, Anne-Marie and {Thomas}, Daniel and {Tremonti}, Christy and {Masters}, Karen and {Coccato}, Lodovico and {Diamond-Stanic}, Aleksandar M. and {Arag{\'o}n-Salamanca}, Alfonso and {Avila-Reese}, Vladimir and {Badenes}, Carles and {Falc{\'o}n-Barroso}, J{\'e}sus and {Belfiore}, Francesco and {Bizyaev}, Dmitry and {Blanc}, Guillermo A. and {Bland-Hawthorn}, Joss and {Blanton}, Michael R. and {Brownstein}, Joel R. and {Byler}, Nell and {Cappellari}, Michele and {Conroy}, Charlie and {Dutton}, Aaron A. and {Emsellem}, Eric and {Etherington}, James and {Frinchaboy}, Peter M. and {Fu}, Hai and {Gunn}, James E. and {Harding}, Paul and {Johnston}, Evelyn J. and {Kauffmann}, Guinevere and {Kinemuchi}, Karen and {Klaene}, Mark A. and {Knapen}, Johan H. and {Leauthaud}, Alexie and {Li}, Cheng and {Lin}, Lihwai and {Maiolino}, Roberto and {Malanushenko}, Viktor and {Malanushenko}, Elena and {Mao}, Shude and {Maraston}, Claudia and {McDermid}, Richard M. and {Merrifield}, Michael R. and {Nichol}, Robert C. and {Oravetz}, Daniel and {Pan}, Kaike and {Parejko}, John K. and {Sanchez}, Sebastian F. and {Schlegel}, David and {Simmons}, Audrey and {Steele}, Oliver and {Steinmetz}, Matthias and {Thanjavur}, Karun and {Thompson}, Benjamin A. and {Tinker}, Jeremy L. and {van den Bosch}, Remco C.~E. and {Westfall}, Kyle B. and {Wilkinson}, David and {Wright}, Shelley and {Xiao}, Ting and {Zhang}, Kai},
        title = "{Overview of the SDSS-IV MaNGA Survey: Mapping nearby Galaxies at Apache Point Observatory}",
      journal = {\apj},
         year = 2015,
        month = jan,
       volume = {798},
       number = {1},
          eid = {7},
        pages = {7},
          doi = {10.1088/0004-637X/798/1/7},
archivePrefix = {arXiv},
       eprint = {1412.1482},
 primaryClass = {astro-ph.GA},
       adsurl = {https://ui.adsabs.harvard.edu/abs/2015ApJ...798....7B}
}

@article{manga2,
   author = {{Yan}, Renbin and {Bundy}, Kevin and {Law}, David R. and {Bershady}, Matthew A. and {Andrews}, Brett and {Cherinka}, Brian and {Diamond-Stanic}, Aleksandar M. and {Drory}, Niv and {MacDonald}, Nicholas and {S{\'a}nchez-Gallego}, Jos{\'e} R. and {Thomas}, Daniel and {Wake}, David A. and {Weijmans}, Anne-Marie and {Westfall}, Kyle B. and {Zhang}, Kai and {Arag{\'o}n-Salamanca}, Alfonso and {Belfiore}, Francesco and {Bizyaev}, Dmitry and {Blanc}, Guillermo A. and {Blanton}, Michael R. and {Brownstein}, Joel and {Cappellari}, Michele and {D'Souza}, Richard and {Emsellem}, Eric and {Fu}, Hai and {Gaulme}, Patrick and {Graham}, Mark T. and {Goddard}, Daniel and {Gunn}, James E. and {Harding}, Paul and {Jones}, Amy and {Kinemuchi}, Karen and {Li}, Cheng and {Li}, Hongyu and {Maiolino}, Roberto and {Mao}, Shude and {Maraston}, Claudia and {Masters}, Karen and {Merrifield}, Michael R. and {Oravetz}, Daniel and {Pan}, Kaike and {Parejko}, John K. and {Sanchez}, Sebastian F. and {Schlegel}, David and {Simmons}, Audrey and {Thanjavur}, Karun and {Tinker}, Jeremy and {Tremonti}, Christy and {van den Bosch}, Remco and {Zheng}, Zheng},
        title = "{SDSS-IV MaNGA IFS Galaxy Survey{\textemdash}Survey Design, Execution, and Initial Data Quality}",
      journal = {\aj},
         year = 2016,
        month = dec,
       volume = {152},
       number = {6},
          eid = {197},
        pages = {197},
          doi = {10.3847/0004-6256/152/6/197},
archivePrefix = {arXiv},
       eprint = {1607.08613},
 primaryClass = {astro-ph.GA},
       adsurl = {https://ui.adsabs.harvard.edu/abs/2016AJ....152..197Y}
}

@article{Leroy2016,
  author = {{Leroy}, Adam K. and {Hughes}, Annie and {Schruba}, Andreas and {Rosolowsky}, Erik and {Blanc}, Guillermo A. and {Bolatto}, Alberto D. and {Colombo}, Dario and {Escala}, Andres and {Kramer}, Carsten and {Kruijssen}, J.~M. Diederik and {Meidt}, Sharon and {Pety}, Jerome and {Querejeta}, Miguel and {Sandstrom}, Karin and {Schinnerer}, Eva and {Sliwa}, Kazimierz and {Usero}, Antonio},
        title = "{A Portrait of Cold Gas in Galaxies at 60 pc Resolution and a Simple Method to Test Hypotheses That Link Small-scale ISM Structure to Galaxy-scale Processes}",
      journal = {\apj},
         year = 2016,
        month = nov,
       volume = {831},
       number = {1},
          eid = {16},
        pages = {16},
          doi = {10.3847/0004-637X/831/1/16},
archivePrefix = {arXiv},
       eprint = {1606.07077},
 primaryClass = {astro-ph.GA},
       adsurl = {https://ui.adsabs.harvard.edu/abs/2016ApJ...831...16L}
}

@article{Sun2018,
  author = {{Sun}, Jiayi and {Leroy}, Adam K. and {Schruba}, Andreas and {Rosolowsky}, Erik and {Hughes}, Annie and {Kruijssen}, J.~M. Diederik and {Meidt}, Sharon and {Schinnerer}, Eva and {Blanc}, Guillermo A. and {Bigiel}, Frank and {Bolatto}, Alberto D. and {Chevance}, M{\'e}lanie and {Groves}, Brent and {Herrera}, Cinthya N. and {Hygate}, Alexander P.~S. and {Pety}, J{\'e}r{\^o}me and {Querejeta}, Miguel and {Usero}, Antonio and {Utomo}, Dyas},
        title = "{Cloud-scale Molecular Gas Properties in 15 Nearby Galaxies}",
      journal = {\apj},
         year = 2018,
        month = jun,
       volume = {860},
       number = {2},
          eid = {172},
        pages = {172},
          doi = {10.3847/1538-4357/aac326},
archivePrefix = {arXiv},
       eprint = {1805.00937},
 primaryClass = {astro-ph.GA},
       adsurl = {https://ui.adsabs.harvard.edu/abs/2018ApJ...860..172S}
}

@ARTICLE{Ellison,
       author = {{Ellison}, Sara L. and {Pan}, Hsi-An and {Bluck}, Asa F.~L. and {Krumholz}, Mark R. and {Lin}, Lihwai and {Hunt}, Leslie and {Corbelli}, Edvige and {Thorp}, Mallory D. and {Barrera-Ballesteros}, Jorge and {S{\'a}nchez}, Sebastian F. and {Scudder}, Jillian M. and {Quai}, Salvatore},
        title = "{The ALMaQUEST Survey XI: a strong but non-linear relationship between star formation and dynamical equilibrium pressure}",
      journal = {\mnras},
         year = 2024,
        month = feb,
       volume = {527},
       number = {4},
        pages = {10201-10220},
          doi = {10.1093/mnras/stad3778},
archivePrefix = {arXiv},
       eprint = {2312.03132},
 primaryClass = {astro-ph.GA},
       adsurl = {https://ui.adsabs.harvard.edu/abs/2024MNRAS.52710201E}
}

@ARTICLE{Lin2019,
       author = {{Lin}, Lihwai and {Pan}, Hsi-An and {Ellison}, Sara L. and {Belfiore}, Francesco and {Shi}, Yong and {S{\'a}nchez}, Sebasti{\'a}n F. and {Hsieh}, Bau-Ching and {Rowlands}, Kate and {Ramya}, S. and {Thorp}, Mallory D. and {Li}, Cheng and {Maiolino}, Roberto},
        title = "{The ALMaQUEST Survey: The Molecular Gas Main Sequence and the Origin of the Star-forming Main Sequence}",
      journal = {\apjl},
         year = 2019,
        month = oct,
       volume = {884},
       number = {2},
          eid = {L33},
        pages = {L33},
          doi = {10.3847/2041-8213/ab4815},
archivePrefix = {arXiv},
       eprint = {1909.11243},
 primaryClass = {astro-ph.GA},
       adsurl = {https://ui.adsabs.harvard.edu/abs/2019ApJ...884L..33L}
}

@ARTICLE{califa,
       author = {{S{\'a}nchez}, S.~F. and {Kennicutt}, R.~C. and {Gil de Paz}, A. and {van de Ven}, G. and {V{\'\i}lchez}, J.~M. and {Wisotzki}, L. and {Walcher}, C.~J. and {Mast}, D. and {Aguerri}, J.~A.~L. and {Albiol-P{\'e}rez}, S. and {Alonso-Herrero}, A. and {Alves}, J. and {Bakos}, J. and {Bart{\'a}kov{\'a}}, T. and {Bland-Hawthorn}, J. and {Boselli}, A. and {Bomans}, D.~J. and {Castillo-Morales}, A. and {Cortijo-Ferrero}, C. and {de Lorenzo-C{\'a}ceres}, A. and {Del Olmo}, A. and {Dettmar}, R. -J. and {D{\'\i}az}, A. and {Ellis}, S. and {Falc{\'o}n-Barroso}, J. and {Flores}, H. and {Gallazzi}, A. and {Garc{\'\i}a-Lorenzo}, B. and {Gonz{\'a}lez Delgado}, R. and {Gruel}, N. and {Haines}, T. and {Hao}, C. and {Husemann}, B. and {Igl{\'e}sias-P{\'a}ramo}, J. and {Jahnke}, K. and {Johnson}, B. and {Jungwiert}, B. and {Kalinova}, V. and {Kehrig}, C. and {Kupko}, D. and {L{\'o}pez-S{\'a}nchez}, {\'A}. R. and {Lyubenova}, M. and {Marino}, R.~A. and {M{\'a}rmol-Queralt{\'o}}, E. and {M{\'a}rquez}, I. and {Masegosa}, J. and {Meidt}, S. and {Mendez-Abreu}, J. and {Monreal-Ibero}, A. and {Montijo}, C. and {Mour{\~a}o}, A.~M. and {Palacios-Navarro}, G. and {Papaderos}, P. and {Pasquali}, A. and {Peletier}, R. and {P{\'e}rez}, E. and {P{\'e}rez}, I. and {Quirrenbach}, A. and {Rela{\~n}o}, M. and {Rosales-Ortega}, F.~F. and {Roth}, M.~M. and {Ruiz-Lara}, T. and {S{\'a}nchez-Bl{\'a}zquez}, P. and {Sengupta}, C. and {Singh}, R. and {Stanishev}, V. and {Trager}, S.~C. and {Vazdekis}, A. and {Viironen}, K. and {Wild}, V. and {Zibetti}, S. and {Ziegler}, B.},
        title = "{CALIFA, the Calar Alto Legacy Integral Field Area survey. I. Survey presentation}",
      journal = {\aap},
         year = 2012,
        month = feb,
       volume = {538},
          eid = {A8},
        pages = {A8},
          doi = {10.1051/0004-6361/201117353},
archivePrefix = {arXiv},
       eprint = {1111.0962},
 primaryClass = {astro-ph.CO},
       adsurl = {https://ui.adsabs.harvard.edu/abs/2012A&A...538A...8S}
}

@ARTICLE{Comerford2024,
       author = {{Comerford}, Julia M. and {Nevin}, Rebecca and {Negus}, James and {Barrows}, R. Scott and {Eracleous}, Michael and {M{\"u}ller-S{\'a}nchez}, Francisco and {Roy}, Namrata and {Stemo}, Aaron and {Storchi-Bergmann}, Thaisa and {Wylezalek}, Dominika},
        title = "{An Excess of Active Galactic Nuclei Triggered by Galaxy Mergers in MaNGA Galaxies of Stellar Mass {\ensuremath{\sim}}{}10$^{11}$ M $_{☉}$}",
      journal = {\apj},
         year = 2024,
        month = mar,
       volume = {963},
       number = {1},
          eid = {53},
        pages = {53},
          doi = {10.3847/1538-4357/ad1a15},
archivePrefix = {arXiv},
       eprint = {2404.14490},
 primaryClass = {astro-ph.GA},
       adsurl = {https://ui.adsabs.harvard.edu/abs/2024ApJ...963...53C}
}

@ARTICLE{Comerford2020,
       author = {{Comerford}, Julia M. and {Negus}, James and {M{\"u}ller-S{\'a}nchez}, Francisco and {Eracleous}, Michael and {Wylezalek}, Dominika and {Storchi-Bergmann}, Thaisa and {Greene}, Jenny E. and {Barrows}, R. Scott and {Nevin}, Rebecca and {Roy}, Namrata and {Stemo}, Aaron},
        title = "{A Catalog of 406 AGNs in MaNGA: A Connection between Radio-mode AGNs and Star Formation Quenching}",
      journal = {\apj},
         year = 2020,
        month = oct,
       volume = {901},
       number = {2},
          eid = {159},
        pages = {159},
          doi = {10.3847/1538-4357/abb2ae},
archivePrefix = {arXiv},
       eprint = {2008.11210},
 primaryClass = {astro-ph.GA},
       adsurl = {https://ui.adsabs.harvard.edu/abs/2020ApJ...901..159C}
}

@ARTICLE{Alban2023,
       author = {{Alb{\'a}n}, M. and {Wylezalek}, D.},
        title = "{Classifying the full SDSS-IV MaNGA Survey using optical diagnostic diagrams: Presentation of AGN catalogs in flexible apertures}",
      journal = {\aap},
         year = 2023,
        month = jun,
       volume = {674},
          eid = {A85},
        pages = {A85},
          doi = {10.1051/0004-6361/202245437},
archivePrefix = {arXiv},
       eprint = {2302.08519},
 primaryClass = {astro-ph.GA},
       adsurl = {https://ui.adsabs.harvard.edu/abs/2023A&A...674A..85A}
}

@ARTICLE{Lacerda2020,
       author = {{Lacerda}, Eduardo A.~D. and {S{\'a}nchez}, Sebasti{\'a}n F. and {Cid Fernandes}, R. and {L{\'o}pez-Cob{\'a}}, Carlos and {Espinosa-Ponce}, Carlos and {Galbany}, L.},
        title = "{Galaxies hosting an active galactic nucleus: a view from the CALIFA survey}",
      journal = {\mnras},
         year = 2020,
        month = mar,
       volume = {492},
       number = {3},
        pages = {3073-3090},
          doi = {10.1093/mnras/staa008},
archivePrefix = {arXiv},
       eprint = {2001.00099},
 primaryClass = {astro-ph.GA},
       adsurl = {https://ui.adsabs.harvard.edu/abs/2020MNRAS.492.3073L}
}

@ARTICLE{MTK06,
       author = {{Mouschovias}, Telemachos Ch. and {Tassis}, Konstantinos and {Kunz}, Matthew W.},
        title = "{Observational Constraints on the Ages of Molecular Clouds and the Star Formation Timescale: Ambipolar-Diffusion-controlled or Turbulence-induced Star Formation?}",
      journal = {\apj},
         year = 2006,
        month = aug,
       volume = {646},
       number = {2},
        pages = {1043-1049},
          doi = {10.1086/500125},
archivePrefix = {arXiv},
       eprint = {astro-ph/0512043},
 primaryClass = {astro-ph},
       adsurl = {https://ui.adsabs.harvard.edu/abs/2006ApJ...646.1043M}
}

@ARTICLE{Kennicutt2007,
       author = {{Kennicutt}, Jr., Robert C. and {Calzetti}, Daniela and {Walter}, Fabian and {Helou}, George and {Hollenbach}, David J. and {Armus}, Lee and {Bendo}, George and {Dale}, Daniel A. and {Draine}, Bruce T. and {Engelbracht}, Charles W. and {Gordon}, Karl D. and {Prescott}, Moire K.~M. and {Regan}, Michael W. and {Thornley}, Michele D. and {Bot}, Caroline and {Brinks}, Elias and {de Blok}, Erwin and {de Mello}, Dulia and {Meyer}, Martin and {Moustakas}, John and {Murphy}, Eric J. and {Sheth}, Kartik and {Smith}, J.~D.~T.},
        title = "{Star Formation in NGC 5194 (M51a). II. The Spatially Resolved Star Formation Law}",
      journal = {\apj},
         year = 2007,
        month = dec,
       volume = {671},
       number = {1},
        pages = {333-348},
          doi = {10.1086/522300},
archivePrefix = {arXiv},
       eprint = {0708.0922},
 primaryClass = {astro-ph},
       adsurl = {https://ui.adsabs.harvard.edu/abs/2007ApJ...671..333K}
}

@ARTICLE{Fukui2021,
       author = {{Fukui}, Yasuo and {Habe}, Asao and {Inoue}, Tsuyoshi and {Enokiya}, Rei and {Tachihara}, Kengo},
        title = "{Cloud-cloud collisions and triggered star formation}",
      journal = {\pasj},
         year = 2021,
        month = jan,
       volume = {73},
        pages = {S1-S34},
          doi = {10.1093/pasj/psaa103},
archivePrefix = {arXiv},
       eprint = {2009.05077},
 primaryClass = {astro-ph.GA},
       adsurl = {https://ui.adsabs.harvard.edu/abs/2021PASJ...73S...1F}
}

@ARTICLE{Reyes2019,
       author = {{de los Reyes}, Mithi A.~C. and {Kennicutt}, Jr., Robert C.},
        title = "{Revisiting the Integrated Star Formation Law. I. Non-starbursting Galaxies}",
      journal = {\apj},
         year = 2019,
        month = feb,
       volume = {872},
       number = {1},
          eid = {16},
        pages = {16},
          doi = {10.3847/1538-4357/aafa82},
archivePrefix = {arXiv},
       eprint = {1901.01283},
 primaryClass = {astro-ph.GA},
       adsurl = {https://ui.adsabs.harvard.edu/abs/2019ApJ...872...16D}
}

@ARTICLE{SanchezGarcia2026,
       author = {{S{\'a}nchez-Garc{\'\i}a}, M. and {D{\'\i}az-Santos}, T. and {Barcos-Mu{\~n}oz}, L. and {Evans}, A.~S. and {Song}, Y. and {Pereira-Santaella}, M. and {Garc{\'\i}a-Burillo}, S. and {Linden}, S.~T. and {Ricci}, C. and {Lenkic}, L. and {Zanella}, A. and {Armus}, L. and {Eibensteiner}, C. and {Teng}, Y.-H. and {Saravia}, A. and {Buiten}, V.~A. and {Privon}, G.~C. and {Torres-Alb{\`a}}, N. and {Saito}, T. and {Larson}, K.~L. and {Bianchin}, M. and {Medling}, A.~M. and {Lai}, T. and {Donnelly}, G.~P. and {Charmandaris}, V. and {Bohn}, T. and {Lofaro}, C.~M. and {Meza}, G.},
        title = "{Spatially resolved star formation relations in local luminous infrared galaxies along the complete merger sequence}",
      journal = {\aap},
         year = 2026,
        month = mar,
       volume = {707},
          eid = {A144},
        pages = {A144},
          doi = {10.1051/0004-6361/202555198},
archivePrefix = {arXiv},
       eprint = {2601.08980},
 primaryClass = {astro-ph.GA},
       adsurl = {https://ui.adsabs.harvard.edu/abs/2026A&A...707A.144S}
}

@ARTICLE{Leroy2021a,
       author = {{Leroy}, Adam K. and {Hughes}, Annie and {Liu}, Daizhong and {Pety}, J{\'e}r{\^o}me and {Rosolowsky}, Erik and {Saito}, Toshiki and {Schinnerer}, Eva and {Schruba}, Andreas and {Usero}, Antonio and {Faesi}, Christopher M. and {Herrera}, Cinthya N. and {Chevance}, M{\'e}lanie and {Hygate}, Alexander P.~S. and {Kepley}, Amanda A. and {Koch}, Eric W. and {Querejeta}, Miguel and {Sliwa}, Kazimierz and {Will}, David and {Wilson}, Christine D. and {Anand}, Gagandeep S. and {Barnes}, Ashley and {Belfiore}, Francesco and {Be{\v{s}}li{\'c}}, Ivana and {Bigiel}, Frank and {Blanc}, Guillermo A. and {Bolatto}, Alberto D. and {Boquien}, M{\`e}d{\`e}ric and {Cao}, Yixian and {Chandar}, Rupali and {Chastenet}, J{\'e}r{\'e}my and {Chiang}, I-Da and {Congiu}, Enrico and {Dale}, Daniel A. and {Deger}, Sinan and {den Brok}, Jakob S. and {Eibensteiner}, Cosima and {Emsellem}, Eric and {Garc{\'\i}a-Rodr{\'\i}guez}, Axel and {Glover}, Simon C.~O. and {Grasha}, Kathryn and {Groves}, Brent and {Henshaw}, Jonathan D. and {Jim{\'e}nez Donaire}, Mar{\'\i}a J. and {Kim}, Jaeyeon and {Klessen}, Ralf S. and {Kreckel}, Kathryn and {Kruijssen}, J.~M. Diederik and {Larson}, Kirsten L. and {Lee}, Janice C. and {Mayker}, Ness and {McElroy}, Rebecca and {Meidt}, Sharon E. and {Mok}, Angus and {Pan}, Hsi-An and {Puschnig}, Johannes and {Razza}, Alessandro and {S{\'a}nchez-Bl'azquez}, Patricia and {Sandstrom}, Karin M. and {Santoro}, Francesco and {Sardone}, Amy and {Scheuermann}, Fabian and {Sun}, Jiayi and {Thilker}, David A. and {Turner}, Jordan A. and {Ubeda}, Leonardo and {Utomo}, Dyas and {Watkins}, Elizabeth J. and {Williams}, Thomas G.},
        title = "{PHANGS-ALMA Data Processing and Pipeline}",
      journal = {\apjs},
         year = 2021,
        month = jul,
       volume = {255},
       number = {1},
          eid = {19},
        pages = {19},
          doi = {10.3847/1538-4365/abec80},
archivePrefix = {arXiv},
       eprint = {2104.07665},
 primaryClass = {astro-ph.IM},
       adsurl = {https://ui.adsabs.harvard.edu/abs/2021ApJS..255...19L}
}

@ARTICLE{Wang2020,
       author = {{Wang}, Tsan-Ming and {Hwang}, Chorng-Yuan},
        title = "{Influence of velocity dispersions on star-formation activities in galaxies}",
      journal = {\aap},
         year = 2020,
        month = sep,
       volume = {641},
          eid = {A24},
        pages = {A24},
          doi = {10.1051/0004-6361/202037748},
       adsurl = {https://ui.adsabs.harvard.edu/abs/2020A&A...641A..24W}
}

\appendix
\nolinenumbers
\clearpage
\section{Classification into group A and B}
\label{app:unsupervised-classification}

The
classification is designed to identify galaxies in which increasing $\Delta v$ is associated with a systematic displacement toward lower $\Sigma_{\rm SFR}$ at approximately fixed $\Sigma_{\rm mol}$. Because this behavior can be confined to only part of the dynamical range of a galaxy, the algorithm is based on identifying it at several resolutions in $\Sigma_{\rm mol}$. The final clustering is fully unsupervised and uses only the three
multiscale diagnostics (features) defined below.

\subsection{Description of the algorithm}
\label{app:classification-inputs}

For every resolved measurement in a galaxy, we define
\begin{equation}
    x \equiv \log\Sigma_{\rm mol},\qquad
    y \equiv \log\Sigma_{\rm SFR},\qquad
    z \equiv \log\Delta v.
    \label{eq:classification-xyz}
\end{equation}
Only measurements for which all three quantities are finite and strictly positive before taking the logarithm are retained. The measurements are
treated independently at this stage; no spatial averaging is introduced by
the classifier.

A galaxy-wide KS relation is first fitted by ordinary least squares,
\begin{equation}
    y = \alpha + \beta x,
    \label{eq:classification-global-ks}
\end{equation}
and the signed vertical residual of each measurement is
\begin{equation}
    r_{\rm KS}
    = y-\left(\alpha+\beta x\right).
    \label{eq:classification-residual}
\end{equation}
Fitting the residuals rather than $y$ directly removes the
galaxy-wide dependence of $\Sigma_{\rm SFR}$ on $\Sigma_{\rm mol}$ before
searching for a $\Delta v$ dependence.

The measurements of each galaxy are sorted by $x$ and divided into
equal-population subsets. For $N$ bins, the bin populations differ by at most
one measurement. The procedure is repeated for $N=3$, 4, and 5.

Within bin $j$ of the $N$-bin partition, the residuals are fitted as
\begin{equation}
    r_{\rm KS}=c_{j,N}+m_{j,N}z.
    \label{eq:classification-local-fit}
\end{equation}
We define the local score
\begin{equation}
    s_{j,N}\equiv -m_{j,N}.
    \label{eq:classification-local-score}
\end{equation}
With this sign convention, a positive score indicates that larger
$\Delta v$ is associated with a more negative KS residual within a restricted
range of $\Sigma_{\rm mol}$. A large positive score therefore represents the
direction of the effect targeted by group~B.

The three galaxy-level features are constructed from the strongest local
scores at each resolution. Let
$s_{(1),N}\geq s_{(2),N}\geq\cdots\geq s_{(N),N}$ denote the scores sorted in
decreasing order. We then define
\begin{align}
    S_3 &= s_{(1),3}, \label{eq:S3}\\
    S_4 &= \frac{1}{2}\left(s_{(1),4}+s_{(2),4}\right), \label{eq:S4}\\
    S_5 &= \frac{1}{3}\left(s_{(1),5}+s_{(2),5}+s_{(3),5}\right).
    \label{eq:S5}
\end{align}
The three-bin feature is therefore maximally sensitive to a strong local
trend, whereas $S_4$ and $S_5$ require the effect to persist over two and
three subsets, respectively. Using all three scales reduces sensitivity to a
single arbitrary binning choice while retaining sensitivity to trends that
are not present across the full $\Sigma_{\rm mol}$ range.

\begin{figure*}
    \centering
    \includegraphics[width=\textwidth]
{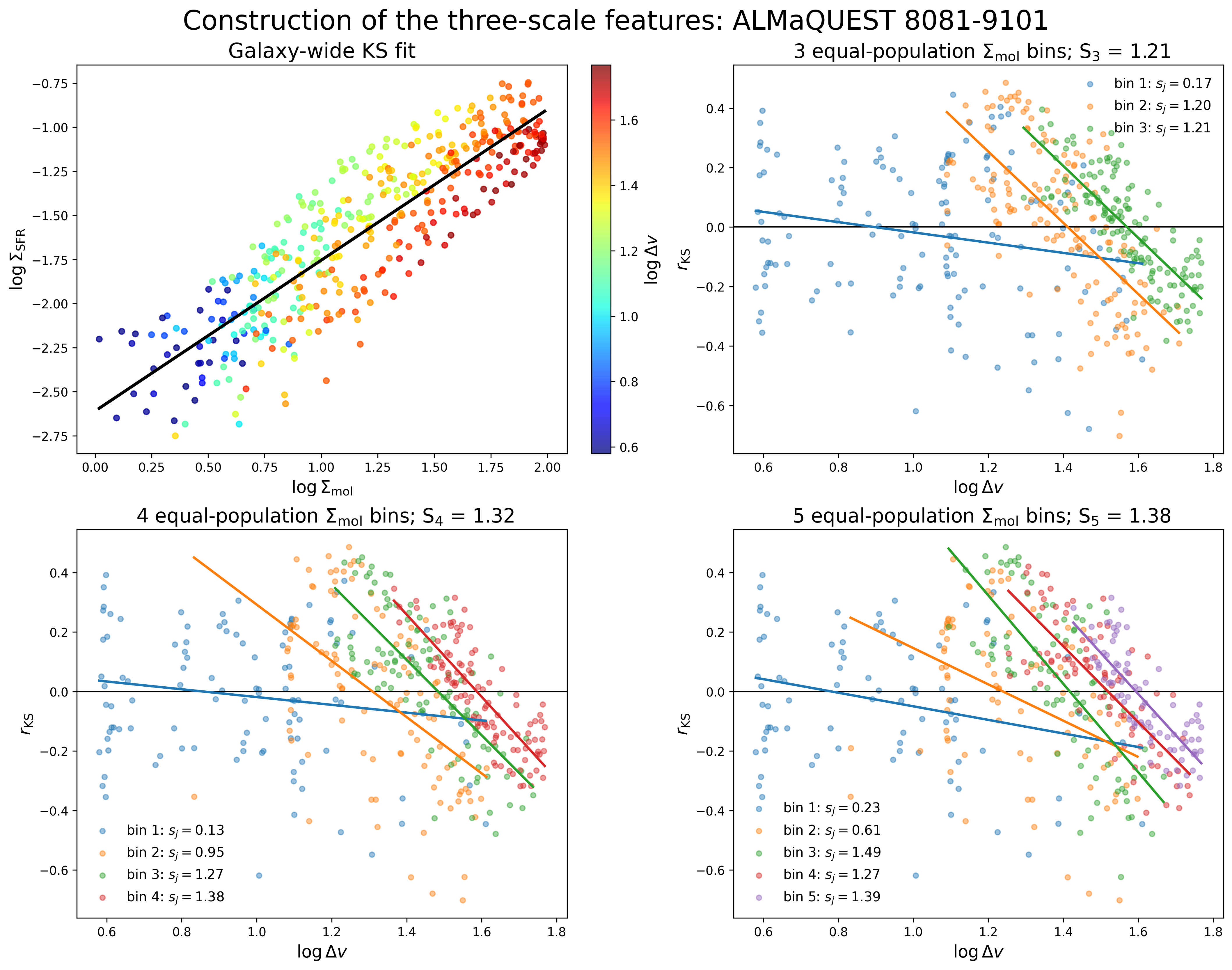}
    \caption{Example of the construction of the three-scale features for
    ALMaQUEST galaxy 8081-9101. The upper-left panel shows the resolved KS
    diagram, colored by $\log\Delta v$, together with the galaxy-wide
    ordinary-least-squares fit. The remaining panels show the signed KS
    residuals against $\log\Delta v$ after division into three, four, and
    five equal-population bins in $\log\Sigma_{\rm mol}$. Each line is
    the local fit in one $\Sigma_{\rm mol}$ bin. The local scores are the negatives
    of the displayed slopes; the strongest one, two, and three scores define
    $S_3$, $S_4$, and $S_5$, respectively.}
    \label{fig:classification-feature-construction}
\end{figure*}

The minimum accepted populations per bin are 15, 12, and 15 measurements for
the three-, four-, and five-bin calculations, respectively. Consequently,
$S_3$, $S_4$, and $S_5$ require at least 45, 48, and 75 valid measurements,
respectively. A feature is also treated as unavailable when the corresponding
bin has no range in $z$.
Figure~\ref{fig:classification-feature-construction} illustrates the full
feature-construction procedure for one galaxy.

\begin{figure*}
    \centering
    \includegraphics[width=\textwidth]
    {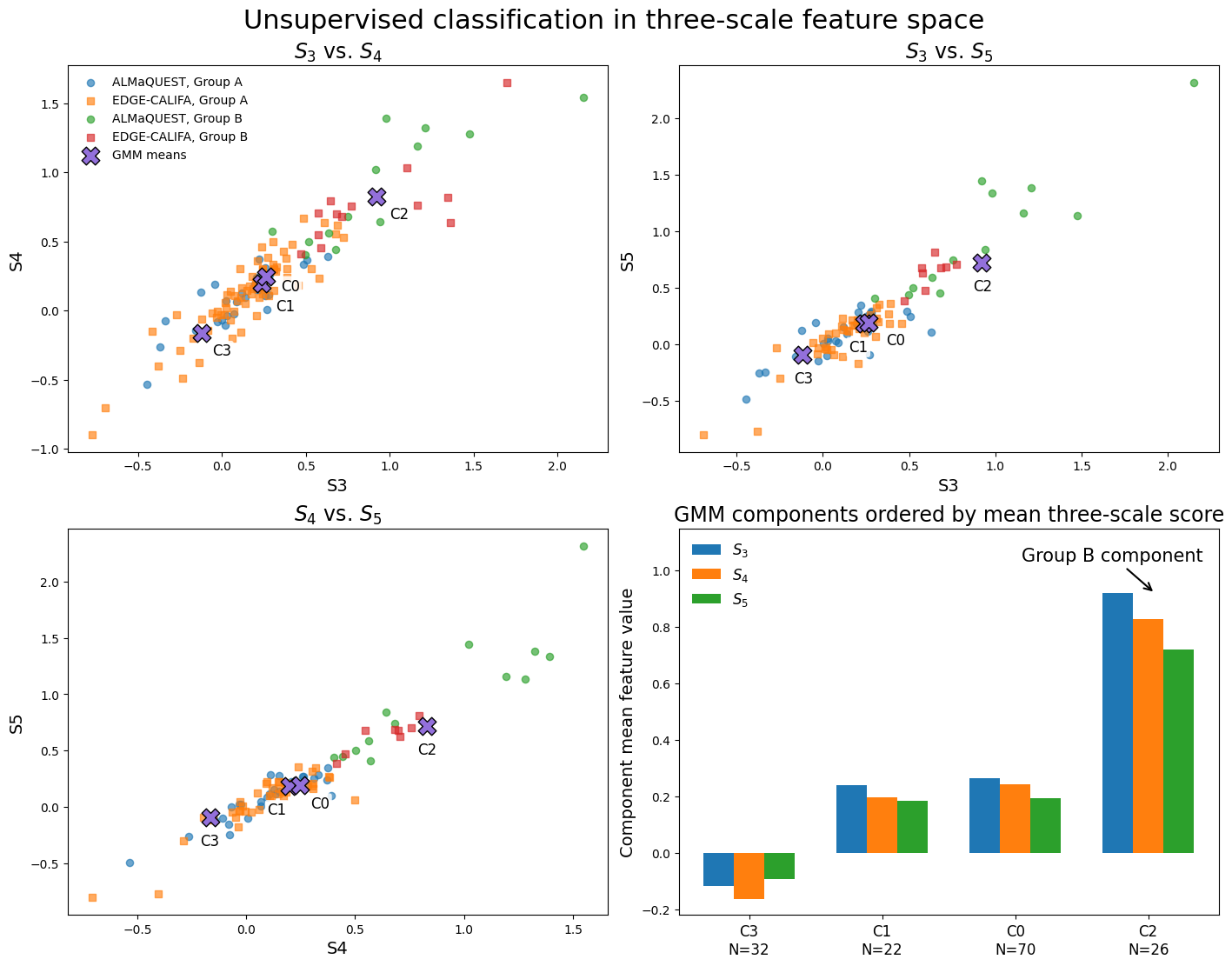}
    \caption{Unsupervised classification in the three-scale feature space.
    The first three panels show pairwise projections of the imputed and
    winsorized $S_3$, $S_4$, and $S_5$ values. Circles and squares denote
    ALMaQUEST and EDGE--CALIFA galaxies, respectively, and the colors denote
    the final groups. Crosses mark the fitted GMM component means. The
    lower-right panel shows the component-mean feature values, with the
    components ordered by increasing mean three-scale score. The
    highest-score component defines group~B; the other three components
    jointly define group~A.}
    \label{fig:classification-feature-space}
\end{figure*}

Each galaxy $i$ is represented by the feature vector
\begin{equation}
    \boldsymbol{S}_i=
    \left(S_{3,i},S_{4,i},S_{5,i}\right).
    \label{eq:classification-feature-vector}
\end{equation}
The ALMaQUEST and EDGE--CALIFA feature vectors are combined before fitting the
clustering model, so that both surveys are classified in one common feature
space.

Some galaxies do not contain enough valid measurements to determine all three features. Missing entries are replaced by the median of the corresponding feature in the combined sample. The number of directly available features is retained as a data-quality diagnostic; galaxies with fewer than three measured features are flagged as low-data cases. The group assignment of galaxies with unavailable features should be interpreted more cautiously because part, or in some cases all, of their position in feature space is determined by median imputation. 
Because group A galaxies constitute the majority of the sample, the median of each feature lies within the region of feature space predominantly occupied by group A. Median imputation therefore tends to place galaxies with unavailable features toward the group A population. This is an intentionally conservative outcome: in the absence of sufficient measurements, the classifier does not infer group B behavior. This should not be interpreted as evidence that group B behavior is intrinsically absent, but rather that the available data are insufficient to establish it reliably.

After imputation, each feature is winsorized at the 1st and 99th percentiles
of the combined sample. Values below or above these limits are replaced by
the corresponding percentile. The winsorized features are then robustly
scaled,
\begin{equation}
    \widetilde S_q=
    \frac{S^{\rm win}_q-\operatorname{median}(S^{\rm win}_q)}
         {\operatorname{IQR}(S^{\rm win}_q)},
    \qquad q\in\{3,4,5\},
    \label{eq:classification-robust-scaling}
\end{equation}
where IQR denotes the interquartile range. Winsorization limits the leverage
of extreme local fits, while robust scaling prevents any one of the three
features from dominating solely because of its numerical range.

The scaled feature vectors are modeled with a four-component Gaussian
mixture,
\begin{equation}
    p(\widetilde{\boldsymbol S})
    =
    \sum_{k=1}^{4}
    \pi_k\,
    \mathcal{N}\!\left(
        \widetilde{\boldsymbol S}\mid
        \boldsymbol{\mu}_k,\sigma_k^2\boldsymbol{I}
    \right),
    \label{eq:classification-gmm}
\end{equation}
where $\pi_k$ is the mixture weight, $\boldsymbol{\mu}_k$ is the component
mean, and each component has a spherical covariance
$\sigma_k^2\boldsymbol{I}$. The fit uses 300 initializations, a maximum of
2000 iterations per initialization, and a fixed random seed of 12345.

The posterior probability that galaxy $i$ belongs to component $k$ is
\begin{equation}
    P_{ik}
    =
    \frac{
        \pi_k\,
        \mathcal{N}\!\left(
            \widetilde{\boldsymbol S}_i\mid
            \boldsymbol{\mu}_k,\sigma_k^2\boldsymbol{I}
        \right)
    }{
        \sum_{\ell=1}^{4}
        \pi_\ell\,
        \mathcal{N}\!\left(
            \widetilde{\boldsymbol S}_i\mid
            \boldsymbol{\mu}_\ell,\sigma_\ell^2\boldsymbol{I}
        \right)
    }.
    \label{eq:classification-posterior}
\end{equation}
The hard component assignment corresponds to the component with the largest posterior
probability.

The four components are ordered using their mean three-scale score. For this
purpose, the imputed and winsorized---but not robustly scaled---features are
used:
\begin{equation}
    \overline{S}_k
    =
    \frac{1}{N_k}
    \sum_{i\in k}
    \frac{
        S^{\rm win}_{3,i}
        +S^{\rm win}_{4,i}
        +S^{\rm win}_{5,i}
    }{3}.
    \label{eq:classification-component-score}
\end{equation}
The component with the largest $\overline{S}_k$ is defined as group~B,
because it contains the galaxies with the strongest and most persistent
negative residual--$\Delta v$ slopes. The other three components collectively
form group~A. The quantity reported as $P(B)$ is the posterior probability
$P_{ik}$ of the group~B component. The classification itself is based on the
maximum-posterior component assignment rather than on a separate
threshold for $P(B)$.

Figure~\ref{fig:classification-feature-space} shows the feature space and the
four fitted components. The use of three group~A components allows the model
to represent the broader structure of the low- and intermediate-score
population without forcing it into a single Gaussian component.

\subsection{Classification catalog}
\label{app:classification-catalog}

The final sample contains 150 galaxies: 45 from ALMaQUEST and 105
from EDGE--CALIFA. The classifier assigns 32 ALMaQUEST galaxies to group~A
and 13 to group~B. For EDGE--CALIFA, 92 galaxies are assigned to group~A and
13 to group~B. Tables~\ref{tab:almaquest_unsupervised_classifications} and~\ref{tab:edge_unsupervised_classifications} give the complete catalog,
including the posterior group~B probability and the three input features.

\begin{table*}[p]
    \centering
    \caption{\textbf{Unsupervised classifications for the ALMaQUEST galaxies.}}
    \label{tab:almaquest_unsupervised_classifications}
    \scriptsize
    \setlength{\tabcolsep}{2.4pt}
    \renewcommand{\arraystretch}{2.0}
    \setlength{\arrayrulewidth}{0.45pt}

    \resizebox{\textwidth}{!}{%
    \begin{tabular}{@{}c@{\hspace{0.75em}\vrule width 0.75pt\hspace{0.22em}\vrule width 0.75pt\hspace{0.75em}}c@{}}
\begin{tabular}[t]{@{}lccccc@{}}
\hline\hline
Galaxy & Group & $P(B)$ & $S_3$ & $S_4$ & $S_5$ \\
\hline
7815-12705 & A & $3.41\times10^{-4}$ & $0.194$ & $0.192$ & $0.181$ \\
7977-12705 & A & $0.0024$ & $-0.123$ & $0.136$ & $0.119$ \\
7977-3703 & A & $1.39\times10^{-4}$ & $-0.025$ & $-0.080$ & $-0.152$ \\
7977-3704 & A & $9.71\times10^{-4}$ & $0.140$ & $0.095$ & $0.088$ \\
8077-6104 & A & $3.86\times10^{-4}$ & $0.290$ & $0.277$ & $0.186$ \\
8077-9101 & A & $2.47\times10^{-4}$ & $0.021$ & $-0.108$ & $-0.100$ \\
8078-12701 & A & $0.0018$ & $0.282$ & $0.258$ & $0.274$ \\
8078-6103 & B & $1.0000$ & $1.473$ & $1.280$ & $1.135$ \\
8081-12703 & B & $1.0000$ & $0.918$ & $1.020$ & $1.446$ \\
8081-3704 & A & $0.0018$ & $0.277$ & $0.261$ & $0.273$ \\
8081-6102 & A & $0.0098$ & $0.630$ & $0.390$ & $0.103$ \\
8081-9101 & B & $1.0000$ & $1.209$ & $1.324$ & $1.382$ \\
8081-9102 & A & $3.79\times10^{-4}$ & $0.222$ & $0.152$ & $0.139$ \\
8082-12701 & A & $0.0017$ & $0.088$ & $0.067$ & $0.009$ \\
8082-12704 & B & $1.0000$ & $0.634$ & $0.563$ & $0.588$ \\
8082-6103 & A & $2.79\times10^{-5}$ & $-0.445$ & $-0.534$ & $-0.490$ \\
8083-12702 & A & $3.28\times10^{-5}$ & $-0.368$ & $-0.263$ & $-0.259$ \\
8083-6101 & B & $1.0000$ & $2.152$ & $1.546$ & $2.315$ \\
8083-9101 & A & $0.0023$ & $0.205$ & $0.151$ & $0.281$ \\
8084-12705 & A & $0.0033$ & $0.505$ & $0.369$ & $0.242$ \\
8084-3702 & A & $0.0017$ & $0.074$ & $-0.025$ & $0.027$ \\
8084-6103 & B & $1.0000$ & $0.752$ & $0.682$ & $0.742$ \\
8086-9101 & A & $5.45\times10^{-4}$ & $0.227$ & $0.207$ & $0.227$ \\
\hline\hline
\end{tabular}
    &
\begin{tabular}[t]{@{}lccccc@{}}
\hline\hline
Galaxy & Group & $P(B)$ & $S_3$ & $S_4$ & $S_5$ \\
\hline
8155-6101 & A & $4.37\times10^{-12}$ & -- & -- & -- \\
8155-6102 & A & $5.50\times10^{-4}$ & $0.260$ & $0.107$ & $0.116$ \\
8156-3701 & A & $0.0010$ & $-0.040$ & $0.193$ & $0.188$ \\
8241-3703 & A & $1.44\times10^{-4}$ & $-0.158$ & $-0.144$ & $-0.114$ \\
8241-3704 & A & $0.0020$ & $0.028$ & $0.069$ & $0.048$ \\
8450-6102 & A & $4.73\times10^{-5}$ & $-0.337$ & $-0.077$ & $-0.247$ \\
8615-12702 & A & $6.28\times10^{-4}$ & $0.268$ & $0.008$ & $-0.097$ \\
8615-3703 & B & $0.9198$ & $0.298$ & $0.571$ & $0.409$ \\
8615-9101 & A & $4.39\times10^{-4}$ & $0.120$ & $0.126$ & $0.153$ \\
8616-12702 & A & $0.0032$ & $0.281$ & $0.112$ & $0.288$ \\
8616-6104 & B & $0.9844$ & $0.499$ & $0.403$ & $0.438$ \\
8616-9102 & A & $0.0096$ & $0.489$ & $0.332$ & $0.290$ \\
8618-9102 & A & $9.64\times10^{-4}$ & $0.001$ & $-0.070$ & $0.001$ \\
8623-12702 & B & $0.9999$ & $0.520$ & $0.501$ & $0.500$ \\
8623-6104 & A & $0.0011$ & $0.259$ & $0.309$ & $0.253$ \\
8655-12705 & B & $1.0000$ & $0.979$ & $1.392$ & $1.337$ \\
8655-3701 & A & $3.44\times10^{-4}$ & $0.206$ & $0.239$ & $0.184$ \\
8655-9102 & A & $0.0016$ & $0.031$ & $-0.034$ & $0.027$ \\
8728-3701 & B & $1.0000$ & $0.941$ & $0.641$ & $0.839$ \\
8950-12705 & B & $1.0000$ & $1.163$ & $1.191$ & $1.158$ \\
8952-12701 & A & $0.0482$ & $0.220$ & $0.373$ & $0.346$ \\
8952-6104 & B & $0.9991$ & $0.676$ & $0.443$ & $0.452$ \\
 \\
\hline\hline
\end{tabular}
    \end{tabular}%
    }

    \par\vspace{0.6ex}
    \begin{minipage}{0.98\textwidth}
        \footnotesize Note. A dash indicates that the
        corresponding feature could not be measured because the relevant
        binning or local fit did not satisfy the data-availability conditions.
    \end{minipage}
\end{table*}

\begin{table*}[p]
    \centering
    \caption{\textbf{Unsupervised classifications for the EDGE--CALIFA galaxies.}}
    \label{tab:edge_unsupervised_classifications}
    \scriptsize
    \setlength{\tabcolsep}{2.4pt}
    \renewcommand{\arraystretch}{1.05}
    \setlength{\arrayrulewidth}{0.45pt}

    \resizebox{\textwidth}{!}{%
    \begin{tabular}{@{}c@{\hspace{0.75em}\vrule width 0.75pt\hspace{0.22em}\vrule width 0.75pt\hspace{0.75em}}c@{}}
\begin{tabular}[t]{@{}lccccc@{}}
\hline\hline
Galaxy & Group & $P(B)$ & $S_3$ & $S_4$ & $S_5$ \\
\hline
ARP220 & A & $0.0164$ & $-0.417$ & $-0.147$ & -- \\
IC0480 & B & $0.9980$ & $1.100$ & $1.034$ & -- \\
IC0540 & A & $4.37\times10^{-12}$ & -- & -- & -- \\
IC0944 & A & $4.40\times10^{-4}$ & $0.199$ & $0.211$ & $0.212$ \\
IC1199 & A & $6.34\times10^{-4}$ & $-0.028$ & $-0.004$ & $-0.033$ \\
IC1683 & A & $0.0140$ & $-0.231$ & $-0.489$ & -- \\
IC2247 & A & $6.31\times10^{-4}$ & $0.026$ & $0.023$ & $-0.046$ \\
IC2487 & A & $9.63\times10^{-4}$ & $0.075$ & $0.107$ & $0.098$ \\
IC4566 & A & $3.82\times10^{-4}$ & $0.177$ & $0.119$ & $0.169$ \\
IC5376 & A & $4.37\times10^{-12}$ & -- & -- & -- \\
NGC0447 & A & $3.39\times10^{-4}$ & $0.249$ & $0.210$ & -- \\
NGC0477 & A & $0.0020$ & $-0.005$ & $-0.027$ & $0.051$ \\
NGC0496 & A & $0.0019$ & $0.274$ & $0.383$ & $0.261$ \\
NGC0523 & A & $8.91\times10^{-5}$ & $-0.378$ & $-0.403$ & $-0.773$ \\
NGC0551 & A & $6.80\times10^{-4}$ & $0.139$ & $0.050$ & $0.124$ \\
NGC1167 & A & $4.72\times10^{-4}$ & $0.210$ & $0.357$ & -- \\
NGC2253 & A & $0.0119$ & $0.309$ & $0.302$ & $0.320$ \\
NGC2347 & A & $5.40\times10^{-4}$ & $0.222$ & $0.093$ & $0.212$ \\
NGC2410 & A & $2.20\times10^{-4}$ & $0.205$ & $-0.038$ & $-0.174$ \\
NGC2480 & A & $4.37\times10^{-12}$ & -- & -- & -- \\
NGC2487 & A & $6.58\times10^{-4}$ & $0.311$ & $0.146$ & $0.228$ \\
NGC2623 & B & $1.0000$ & $1.698$ & $1.649$ & -- \\
NGC2639 & A & $8.77\times10^{-5}$ & $-0.693$ & $-0.706$ & $-0.802$ \\
NGC2730 & A & $0.0012$ & $0.033$ & $0.113$ & $0.093$ \\
NGC2906 & A & $0.0476$ & $0.325$ & $0.317$ & $0.348$ \\
NGC2916 & B & $0.9987$ & $1.349$ & $0.820$ & -- \\
NGC3303 & A & $4.37\times10^{-12}$ & -- & -- & -- \\
NGC3381 & A & $4.37\times10^{-12}$ & -- & -- & -- \\
NGC3811 & A & $5.74\times10^{-4}$ & $0.251$ & $0.150$ & $0.226$ \\
NGC3815 & A & $0.0214$ & $0.484$ & $0.667$ & -- \\
NGC3994 & A & $7.26\times10^{-4}$ & $0.456$ & $0.182$ & $0.182$ \\
NGC4047 & A & $5.44\times10^{-4}$ & $0.150$ & $0.147$ & $0.114$ \\
NGC4149 & A & $0.0017$ & $0.419$ & $0.480$ & -- \\
NGC4185 & A & $4.37\times10^{-12}$ & -- & -- & -- \\
NGC4210 & A & $3.76\times10^{-4}$ & $0.239$ & $0.197$ & $0.198$ \\
NGC4211NED02 & A & $4.37\times10^{-12}$ & -- & -- & -- \\
NGC4470 & A & $4.28\times10^{-4}$ & $-0.270$ & $-0.028$ & $-0.033$ \\
NGC4644 & A & $4.58\times10^{-4}$ & $0.111$ & $0.306$ & $0.161$ \\
NGC4676A & A & $4.37\times10^{-12}$ & -- & -- & -- \\
NGC4711 & A & $3.81\times10^{-4}$ & $0.208$ & $0.182$ & $0.134$ \\
NGC4961 & A & $0.0053$ & $-0.122$ & $-0.064$ & -- \\
NGC5000 & A & $0.0571$ & $0.725$ & $0.529$ & -- \\
NGC5016 & B & $1.0000$ & $0.575$ & $0.706$ & $0.629$ \\
NGC5056 & A & $4.64\times10^{-4}$ & $0.321$ & $0.284$ & $0.199$ \\
NGC5205 & A & $4.19\times10^{-4}$ & $0.310$ & $0.299$ & -- \\
NGC5218 & A & $0.0806$ & $0.390$ & $0.239$ & $0.357$ \\
NGC5394 & A & $0.0913$ & $0.688$ & $0.621$ & -- \\
NGC5406 & B & $1.0000$ & $0.575$ & $0.547$ & $0.679$ \\
NGC5480 & A & $0.0031$ & $0.382$ & $0.380$ & $0.268$ \\
NGC5520 & B & $1.0000$ & $0.716$ & $0.683$ & $0.683$ \\
NGC5614 & A & $0.0047$ & $0.304$ & $0.497$ & $0.067$ \\
NGC5633 & B & $0.5700$ & $0.473$ & $0.413$ & $0.384$ \\
NGC5657 & A & $4.37\times10^{-12}$ & -- & -- & -- \\
\hline\hline
\end{tabular}
    &
\begin{tabular}[t]{@{}lccccc@{}}
\hline\hline
Galaxy & Group & $P(B)$ & $S_3$ & $S_4$ & $S_5$ \\
\hline
NGC5732 & A & $4.37\times10^{-12}$ & -- & -- & -- \\
NGC5784 & A & $0.0137$ & $-0.172$ & $-0.201$ & -- \\
NGC5908 & A & $6.75\times10^{-4}$ & $0.242$ & $0.172$ & $0.095$ \\
NGC5930 & A & $3.41\times10^{-4}$ & $0.205$ & -- & -- \\
NGC5934 & A & $0.0413$ & $0.679$ & $0.558$ & -- \\
NGC5947 & A & $4.37\times10^{-12}$ & -- & -- & -- \\
NGC5953 & A & $2.51\times10^{-4}$ & $0.115$ & $-0.154$ & $-0.111$ \\
NGC5980 & A & $3.47\times10^{-4}$ & $0.185$ & $0.157$ & $0.165$ \\
NGC6004 & A & $5.98\times10^{-4}$ & $0.014$ & $-0.033$ & $-0.039$ \\
NGC6060 & A & $5.55\times10^{-4}$ & $0.387$ & $0.301$ & $0.184$ \\
NGC6155 & A & $8.05\times10^{-4}$ & $0.112$ & $0.094$ & $0.226$ \\
NGC6168 & B & $0.9931$ & $1.164$ & $0.763$ & -- \\
NGC6186 & A & $9.27\times10^{-4}$ & $0.020$ & $0.060$ & $-0.024$ \\
NGC6301 & B & $0.9995$ & $0.592$ & $0.456$ & $0.472$ \\
NGC6310 & A & $4.37\times10^{-12}$ & -- & -- & -- \\
NGC6314 & A & $0.0160$ & $-0.137$ & $-0.374$ & -- \\
NGC6361 & A & $4.75\times10^{-4}$ & $0.244$ & $0.307$ & $0.207$ \\
NGC6394 & A & $0.0072$ & $-0.086$ & $-0.144$ & -- \\
NGC6478 & A & $4.97\times10^{-4}$ & $0.168$ & $0.176$ & $0.216$ \\
NGC7738 & A & $0.0127$ & $-0.772$ & $-0.897$ & -- \\
NGC7819 & A & $6.01\times10^{-4}$ & $0.094$ & $0.068$ & -- \\
UGC00809 & A & $4.37\times10^{-12}$ & -- & -- & -- \\
UGC03253 & A & $4.37\times10^{-12}$ & -- & -- & -- \\
UGC03539 & A & $0.0020$ & $0.577$ & $0.233$ & -- \\
UGC03969 & A & $9.00\times10^{-4}$ & $0.236$ & $0.463$ & -- \\
UGC03973 & A & $2.58\times10^{-4}$ & $0.063$ & $-0.197$ & $-0.091$ \\
UGC04029 & B & $1.0000$ & $0.682$ & $0.699$ & $0.677$ \\
UGC04132 & A & $4.82\times10^{-4}$ & $0.120$ & $0.167$ & $0.129$ \\
UGC04280 & A & $4.37\times10^{-12}$ & -- & -- & -- \\
UGC04461 & A & $3.54\times10^{-4}$ & $0.182$ & $0.244$ & -- \\
UGC05108 & A & $4.37\times10^{-12}$ & -- & -- & -- \\
UGC05111 & A & $0.0011$ & $-0.059$ & $-0.015$ & $0.011$ \\
UGC05359 & A & $0.0015$ & $0.534$ & $0.301$ & -- \\
UGC05598 & A & $4.37\times10^{-12}$ & -- & -- & -- \\
UGC07012 & A & $4.37\times10^{-12}$ & -- & -- & -- \\
UGC08107 & A & $2.99\times10^{-5}$ & $-0.249$ & $-0.286$ & $-0.301$ \\
UGC08267 & A & $4.11\times10^{-4}$ & $0.281$ & $0.110$ & -- \\
UGC09067 & A & $9.67\times10^{-4}$ & $0.073$ & $-0.003$ & -- \\
UGC09476 & A & $5.53\times10^{-4}$ & $0.049$ & $-0.066$ & $-0.046$ \\
UGC09537 & A & $2.84\times10^{-4}$ & $-0.033$ & $-0.048$ & $-0.088$ \\
UGC09542 & A & $5.89\times10^{-4}$ & $0.049$ & $0.138$ & -- \\
UGC09665 & B & $1.0000$ & $0.773$ & $0.759$ & $0.705$ \\
UGC09759 & B & $0.9969$ & $1.361$ & $0.638$ & -- \\
UGC09873 & A & $4.37\times10^{-12}$ & -- & -- & -- \\
UGC09892 & A & $3.83\times10^{-4}$ & $0.245$ & $0.121$ & -- \\
UGC09919 & A & $4.37\times10^{-12}$ & -- & -- & -- \\
UGC10043 & A & $0.0444$ & $0.608$ & $0.637$ & -- \\
UGC10123 & B & $1.0000$ & $0.647$ & $0.792$ & $0.813$ \\
UGC10205 & A & $4.66\times10^{-4}$ & $0.306$ & $0.335$ & -- \\
UGC10380 & A & $4.37\times10^{-12}$ & -- & -- & -- \\
UGC10384 & A & $3.78\times10^{-4}$ & $0.245$ & $0.296$ & -- \\
UGC10710 & A & $9.11\times10^{-4}$ & $0.365$ & $0.427$ & -- \\
 &  & -- & -- & -- & -- \\
\hline\hline
\end{tabular}
    \end{tabular}%
    }

    \par\vspace{0.6ex}
    \begin{minipage}{0.98\textwidth}
        \footnotesize Note. 
        A dash indicates that the
        corresponding feature could not be measured because the relevant
        binning or local fit did not satisfy the data-availability conditions.
    \end{minipage}
\end{table*}

\section{Sampling limitations of the PHANGS--ALMA velocity-dispersion maps}
\label{app:phangs}

\begin{figure}
  \includegraphics[width=\hsize]{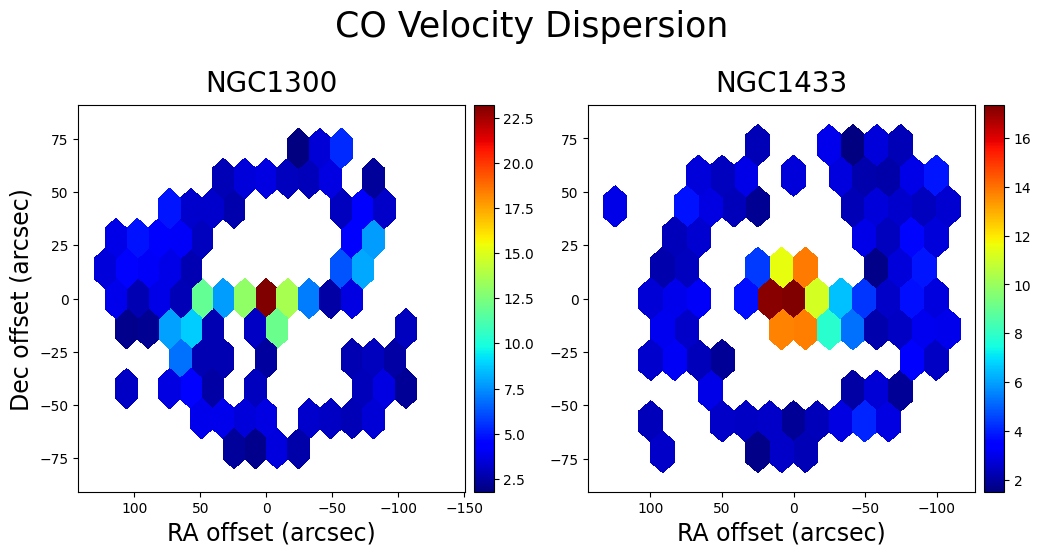}
  \caption{Velocity dispersion ($\Delta v$) maps of two typical PHANGS--ALMA galaxies. They are sparsely sampled, and regions close to the centers are removed.}
  \label{phangs_example_maps_Dv}
\end{figure}

Measurements of higher-order moments, such as the velocity dispersion, are particularly sensitive to noise. The PHANGS--ALMA linewidth maps are therefore constructed using a high-confidence "strict" emission mask, which prioritizes the reliability of the retained measurements over spatial completeness and excludes faint, low signal-to-noise emission \citep{Leroy2021a}. Consequently, the available $\Delta v$ maps are often sparsely sampled. We present example maps in Fig.~\ref{phangs_example_maps_Dv}. We further note that the missing sightlines often include substantial parts of the inner regions of the galaxies, thereby reducing the coverage of regions containing some of the largest measured values of $\Delta v$ (see Fig.~\ref{alma_cleaning_example}). The resulting loss of spatial coverage and $\Delta v$ dynamical range reduces the sensitivity of our classification to a component of the $\Delta v$ gradient perpendicular to the KS relation. We therefore interpret the absence of group~B galaxies in PHANGS--ALMA as being at least partly related to the conservative construction and sparse sampling of its linewidth products, rather than as conclusive evidence that such behavior is intrinsically absent from the PHANGS--ALMA galaxy population.

\section{Dependence on galaxy inclination}
\label{app:inclination}

\begin{figure*}[!b]
  \includegraphics[width=\hsize]{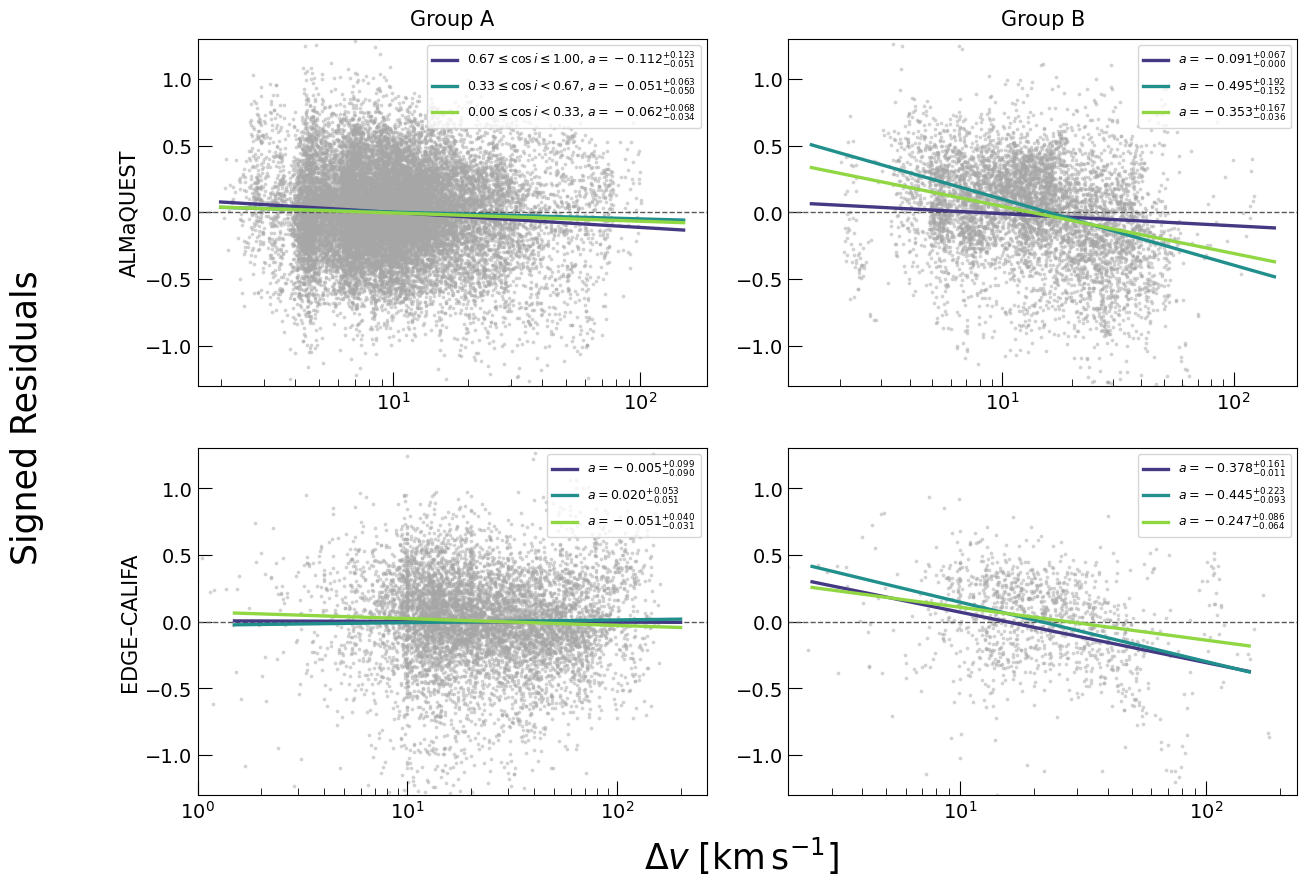}
  \caption{$Res-\Delta v$ relation within three inclination bins. There is no systematic trend with inclination.}
  \label{fig:inclination_bins}
  \vspace{9cm}
\end{figure*}

To investigate the effect of inclination angle on the $Res-\Delta v$ trend, we divide the galaxies in each survey into three equal bins according to $\cos i$, and, for each group separately, we repeat the analysis described in Section~3 (i.e., fitting the KS relation and then $Res-\log\Delta v$) within each inclination bin. The fitted $Res$--$\Delta v$ slopes show no systematic dependence on $\cos i$ (Fig.~\ref{fig:inclination_bins}). Thus, within the range of inclinations and spatial resolutions sampled by these datasets, we find no evidence that the observed trend of decreasing $Res$ with increasing $\Delta v$ is driven by galaxy inclination. 
We note that this test does not exclude small inclination-dependent contributions to the measured linewidths, nor does it fully isolate beam smearing or unresolved rotation within individual galaxies. Nevertheless, the absence of a systematic trend across the inclination bins indicates that inclination alone does not account for the difference between the two groups.

\end{document}